\documentclass{aa}
\usepackage{amsmath}
\usepackage{graphicx}
\usepackage[caption=false]{subfig}
\usepackage{appendix}
\usepackage{threeparttable}
\usepackage{txfonts}
\usepackage{color}
\usepackage{float}
\usepackage[version=4]{mhchem} 
\usepackage{booktabs}
\usepackage{placeins}
\usepackage{balance}

\usepackage{natbib}
\bibpunct{(}{)}{;}{a}{}{,} 

\usepackage{epstopdf}

\usepackage[off]{auto-pst-pdf}
\begin{document}

\title{Laboratory spectroscopy of theoretical ices: Predictions for JWST and test for astrochemical models \thanks{Based on experiments conducted at the Max-Planck-Institut f\"ur Extraterrestrische Physik, Garching.},\thanks{The raw spectra are available in electronic form at the CDS via anonymous ftp to cdsarc.u-strasbg.fr (<...>) or via http://cdsweb.u-strasbg.fr/<...>}}

\author{B. M\"uller \inst{\ref{MPE}}
        \and
        B. M. Giuliano \inst{\ref{MPE}}
        \and
        A. Vasyunin \inst{\ref{UrFU}}
        \and
        G. Fedoseev \inst{\ref{UrFU}}
        \and
        P. Caselli \inst{\ref{MPE}}}

\offprints{Birgitta M\"uller}

\institute{Max-Planck-Institut f\"ur extraterrestrische Physik, Gie{\ss}enbachstra{\ss}e 1, 85748 Garching bei M\"unchen, Germany \label{MPE}\\
       \email{bmueller@mpe.mpg.de}
           \and
           Research Laboratory for Astrochemistry, Ural Federal University, 620002, 19 Mira street, Yekaterinburg, Russia \label{UrFU}}

\date{Received 2 February 2022 / Accepted 29 July 2022}
\authorrunning{B. M\"uller et al.}  
\titlerunning{Layers}

\abstract
{The pre-stellar core L1544 has been the subject of several observations conducted in the past years, complemented by modelling studies focused on its gas and ice-grain chemistry. The chemical composition of the ice mantles reflects the environmental physical changes along the temporal evolution, such as density and temperature. The investigation outcome hints at a layered structure of interstellar ices with abundance of \ce{H2O} in the inner layers and an increasing concentration of CO near the surface. The morphology of interstellar ice analogues can be investigated experimentally assuming a composition derived from chemical models.}
{This research presents a new approach of a three-dimensional fit where observational results are first fitted with a gas-grain chemical model predicting the exact ice composition including infrared (IR) inactive species. Then the laboratory IR spectra are recorded for interstellar ice analogues whose compositions reflect the obtained numerical results, in a layered and in a mixed morphology. These results could then be compared with the results of James Webb Space Telescope (JWST) observations.
Special attention is paid to the inclusion of IR inactive species whose presence is predicted in the ice, but is typically omitted in the laboratory obtained data. This stands for \ce{N2}, one of the main possible constituents of interstellar ice mantles, and \ce{O2}.}
{Ice analogue spectra were recorded at a temperature of 10~K using a Fourier transform infrared spectrometer. 
In the case of layered ice we deposited a \ce{H2O}-CO-\ce{N2}-\ce{O2} mixture on top of a \ce{H2O}-\ce{CH3OH}-\ce{N2} ice, while in the case of mixed ice we examined a \ce{H2O}-\ce{CH3OH}-\ce{N2}-CO composition. The selected species are the four most abundant ice components predicted by the chemical model.}  
{Following the changing composition and structure of the ice, we find differences in the absorption bands for most of the examined vibrational modes. The extent of observed changes in the IR band profiles will allow us  to analyse the structure of ice mantles in L1544 from future observations by the JWST.}
{Our spectroscopic measurements of interstellar ice analogues predicted by our well-received gas-grain chemical codes of pre-stellar cores will allow detailed comparison with upcoming JWST observations. This is crucial in order to put stringent constraints on the chemical and physical structure of dust icy mantles just before the formation of stars and protoplanetary disks, and  to explain  surface chemistry.}


\maketitle

\section{Introduction}
\label{intro}
\indent\indent The James Webb Space Telescope (JWST) is the next-generation National Aeronautics and Space Administration (NASA) telescope which is designed to work with enhanced sensitivity covering the infrared spectroscopic range. The applications of its capabilities are many, among them the investigation of ice mantles in different astronomical environments such as star-forming regions from the earliest phases, represented by pre-stellar cores, protostars, and planet-forming disks.

In preparation of upcoming JWST observations, preliminary work is required to simulate the conditions of the astronomical object of interest and acquire laboratory spectral data which will facilitate and reinforce the observational data analysis once they become available.
The interpretation of the observational data from telescope facilities can be greatly assisted by theoretical modelling and laboratory experiments.
In this paper we present our custom-designed experiments, guided by chemical simulations on the pre-stellar core L1544 using a gas-grain chemical model. The aim of the work is to make available to the community a set of laboratory spectra that mimic the composition of ice mantles on top of the dust grains predicted by the results of theoretical simulations presented in \citet{Vasyunin2017}; these spectra will  be available for comparison with the future data from JWST.

L1544 is a  typical pre-stellar core, with clear signs of contraction motions \citep[e.g.][]{Keto2010,Caselli2012} and a highly concentrated structure where the density goes up to about 10$^7$ \ce{H2} molecules~cm$^{-3}$ and temperatures drop to about 6~K \citep{Crapsi2007, Keto2015}. Recent Atacama Large Millimeter/submillimeter Array (ALMA) observations have detected a compact central region with radius $\simeq$ 1400~au, called the kernel \citep{Caselli2019}, where almost all species heavier than He have been found to freeze out onto dust grains \citep{Caselli2022}, as predicted by chemical models \citep{Vasyunin2017,Sipila2019}.

By exploring different ice chemical  compositions and thicknesses in the laboratory, we offer a tool to disentangle the contribution to the spectroscopic fingerprints from the internal region of the cloud, the kernel, from the entire line of sight. Provided the presence of a background star that is luminous enough to shine through the denser region with high extinction, called the dust peak, and still be detected by the telescope, it will be therefore possible to utilise observational data to probe the ice mantle compositions along the cloud depth.

We   previously investigated the spectroscopic signature of ice analogues in \citet{Mueller2018} and \citet{Mueller2021}. In these studies, the chemical composition and physical state were adjusted to the available results from observational data (see references therein).

The main novelty of this research is the presentation of the new approach when observational results are first fitted with the gas-grain chemical model from \citet{Vasyunin2017} predicting the exact ice composition including infrared (IR) inactive species. Then the laboratory IR spectra are recorded for the interstellar ice analogues whose composition reflects the obtained numerical results. These results could then be compared with the results of JWST observations. In this way we provide the first attempt to execute a three-dimensional fit  between the observations, models and laboratory work.
Special attention is paid to the inclusion of IR inactive species whose presence is predicted in the ice but is mostly omitted in the laboratory obtained data (cf. Leiden Ice Database\footnote{https://icedb.strw.leidenuniv.nl/}, for example),  such as \ce{N2}, one of the main possible constituents of the interstellar ice mantles, and \ce{O2}.
Although a rich variety of spectroscopic data of ice mantle analogues are already accessible from the commonly known databases, data on the desired composition and structure predicted by chemical models can be retrieved only with a dedicated experiment.
We provide laboratory results for both a layered and a mixed morphology of ice analogues to  reflect the inhomogeneity of the realistic ice mantles with respect to those typically obtained in the lab.

In the next sections we  present a description of the model employed and its predictions (Section~\ref{model}), followed by an overview of the experimental methodology in Section~\ref{setup}. The discussion of the experimental results is conducted in Section~\ref{results} and Section~\ref{discuss}, followed by our conclusions in Section~\ref{concl}. Additionally, a comparison of the computationally added layer spectrum with the experimentally layered ices is addressed in  Appendix~\ref{compar_l_meas_comp}.

\section{Model description}
\label{model}
\indent\indent The theoretical icy mantles are those obtained by \citet{Vasyunin2017} and we report here some basic descriptions of the model, avoiding details that can be found there.  \citet{Vasyunin2017} applied their three-phase chemical model to the physical structure (volume density, dust and gas temperature, velocity profiles) of L1544 as described in \citet{Keto2015}. A three-phase model indicates a gas-grain chemical model where three different environments are taken into account: (i) the gas-phase; (ii) the surface of icy mantles (represented by the first four monolayers, see~\citealt[][]{Vasyunin2013}); and (iii) the bulk of icy mantles (all monolayers below the surface). The model provides a level of detail high enough to be comparable with the observational data, while also being fast. Rate equations, with modifications to take into account stochastic effects, are used. Modifications are particularly important, as hydrogen atoms and molecules (H and \ce{H2}) are assumed to diffuse across the icy mantles, though in our case it occurs mainly through quantum-tunnelling \citep[see e.g.][]{Garrod2009}. Dust grains are assumed to be spherical, with sizes equal to 0.1\,{$\mu$}m.

Five desorption processes are included in the model: thermal desorption, cosmic-ray induced desorption, cosmic-ray induced photodesorption, photodesorption, and reactive desorption, with only surface species allowed to desorb. Although the first four desorption mechanisms were implemented using standard procedures \citep[e.g.][]{Hasegawa1993,Prasad1983}, \citet{Vasyunin2017} implemented new experimental work from \citet{Minissale2016} to follow the reactive desorption during the chemical evolution of the pre-stellar core. The most important experimental finding of \citet{Minissale2016} was that the reactive desorption efficiency of a molecule formed onto the surface of a dust grain is a sensitive function of the surface composition, with CO-rich (\ce{H2O}-rich) ices allowing the most (least) efficient reactive desorption. This is especially important for methanol, which mainly forms on the surface of dust grains via successive hydrogenation of CO. Within the catastrophic CO freeze-out zone of pre-stellar cores \citep[which in L1544 starts at a radius of $\simeq$7000\,au;][]{Caselli_1999}, where more than 90~\% of CO molecules are adsorbed onto dust grains, CO-rich icy mantles allow efficient surface formation and reactive desorption of \ce{CH3OH}, in good agreement with observations \citep[see also][for detailed comparisons of the model predictions with observations of other complex organic molecules within L1544]{Jimenez-Serra2016}.

The methanol peak observed towards the L1544 pre-stellar core could successfully be reproduced with the original model by \citet{Vasyunin2017}. Since then the model has been applied to several other cores. Updates made to the original model are described in \citet{Nagy2019}, \citet{Lattanzi2020}, \citet{Jimenez-Serra2021}, and \citet{Scibelli2021}.
This comprehensive gas-grain chemical network, applied to the L1544 physical structure, provides the chemical composition of the icy mantles at each radius of the pre-stellar core. An example of this output is shown in Fig.~\ref{L1544_dust_peak}, where the fractional abundances of the most abundant solid species within an icy mantle of a dust grain located 1160\,au  from the L1544 centre (representative of the icy mantle composition within the L1544 kernel), are plotted as a function of the number of monolayers. The \ce{O2} fraction in the outer layers of the icy mantles is in agreement with \textit{\emph{in situ}} measurements of \ce{O2} in comet 67P \citep{Bieler2015}. The large increase in \ce{O2} in the outer layers of the icy mantles is due to the fact that the O/H ratio increases towards the central regions of pre-stellar cores, so that \ce{O2} is formed more readily than \ce{H2O}. Figure \ref{L1544_dust_peak} is discussed in the next section.

\section{Experimental methods}
\label{setup}
\indent\indent All the experiments presented in this paper were conducted at the Max Planck Institute for Extraterrestrial Physics in Garching (Germany), using the cryogenic set-up developed at the Center for Astrochemical Studies (CAS). A full description of the set-up can be found in \citet{Mueller2018} and \citet{Mueller2021}; here we describe the most relevant details for the presentation of our results.

\subsection{Cryogenic set-up}

The set-up is composed of a closed-cycle He cryostat (Advanced Research Systems) with a base pressure of $10^{-5}$ mbar at ambient temperature, coupled with a Bruker Fourier transform infrared (FTIR) spectrometer. The sample compartment of the spectrometer hosts the cryostat mounted in a stainless steel vacuum chamber, which is able to achieve a minimum temperature of 10~K and a final vacuum of $10^{-7}$ mbar when cold. The vacuum chamber is equipped with ZnSe optical windows, while KBr is used as cold substrate window.

The IR spectra were recorded in the 4800--500~cm$^{-1}$ (2.1--20~$\mu$m) frequency range using a standard deuterated triglycine sulfate (DTGS) detector. A spectral resolution of 1~cm$^{-1}$ was used and the signal was averaged over 128 scans.

The ice formation is attained by condensation of a suitable gaseous mixture on top of the cold substrate. The following gases were employed in these experiments: water vapour from double distilled water; methanol vapour from a liquid sample by Sigma Aldrich with 99.8~\% purity; and CO, \ce{N2}, and \ce{O2} from gas bottles with over 99~\% purity.
Gases were pre-mixed in a glass bulb in relative proportions controlled by measuring the partial pressure of each species, following the ideal gas law. In the case of the layered ice analogue, two different gas mixtures were expanded separately into the vacuum chamber by using a two-fold gas inlet, to allow the formation of a layered structure.
Deposition rates for the layered ice were 0.14 $\mu$m s$^{-1}$ for layer 1 and 0.04 $\mu$m s$^{-1}$ for layer 2 to ensure similar deposition times for the two different ice thicknesses. Layer 2 was deposited immediately after layer 1. For the mixed ice, a similar deposition rate as for layer 2 was used. We observed no impact on the ice structures due to differences in the deposition rates.
No contamination from residual gas in the inlet pipe was observed within the sensitivity of our techniques.

The relative abundances of the molecular species in the ice samples, after deposition, were estimated by the integrated absorption features calculating the column density $N$ from the corresponding band strengths $A$ as
\begin{equation}
N = \mathrm{ln}(10) \cdot \frac{Area}{A},
\end{equation}
where $Area$ is the area in absorbance of the corresponding absorption band. Band strength values were taken from the literature and are listed in Table \ref{mol_band_str}.
From the column density (molecules~cm$^{-2}$), the ice thickness $d$ (cm) could then be determined using
\begin{equation}
    d = \frac{N \cdot M}{\rho \cdot N_A},
\end{equation}
where $M$ is the molecular mass (g~mol$^{-1}$), $\rho$ the density (g~cm$^{-3}$), and $N_A$ the Avogadro constant.
We note that all band strength values were taken from measurements of pure ices or ice matrices that do not fit the presented experimental conditions perfectly. This contributes to an error in the calculation of the column density and ultimately the ice thickness, as noted in Table \ref{exps_real}.
For the density of all ice analogues, we assumed $\rho$ = 0.8~g~cm$^{-3}$ as this is the density of amorphous solid water, a major component of our ice matrices, when deposited in a 45$^{\circ}$ angle onto the substrate \citep{Dohnalek2003}.
Assuming that one monolayer has a thickness of approximately 3 \AA~ \citep[cf.][]{Opitz2007, Potapov2020}, the thickness of our ice analogues can be expressed in terms of monolayers (MLs) to fit the terms of the theoretical model, as noted in Table \ref{exps_real}.

\begin{table*}[!htp]
\centering
\begin{threeparttable}
\caption{Band strength $A$, ice composition, temperature $T$, and references of species used in the presented experiments.}
\label{mol_band_str}
\begin{tabular}{c || l l l c c}
\toprule \toprule
Molecule   & Vibration mode                    & $A$ (cm molec$^{-1}$) & composition  & $T$ (K) & Reference \\ [0.5ex]
\midrule
\ce{H2O}   & $\nu_1$, $\nu_3$ OH stretch       & 2.0 $\times 10^{-16}$ & pure         & 10    & $a$             \\ [0.5ex]
           & $\nu_2$ OH bend                       & 1.2 $\times 10^{-17}$ & pure         & 14    & $b$ \\ [0.5ex]
\ce{CH3OH} & $\nu_9$ CH stretch                    & 2.1 $\times 10^{-17}$ & pure         & 10    & $c$ \\ [0.5ex]
               & $\nu_3$ CH stretch                & 5.3 $\times 10^{-18}$ & pure         & 10    & $c$ \\ [0.5ex]
               & \ce{CH3} bend                         & 1.2 $\times 10^{-17}$ & pure         & 10    & $c$ \\ [0.5ex]
               & $\nu_8$ CO stretch                        & 1.8 $\times 10^{-17}$ & pure         & 10    & $d$ \\ [0.5ex]
CO             & CO stretch                            & 1.1 $\times 10^{-17}$ & pure         & 30    & $e$ \\ [0.5ex]
\ce{N2}    &                                               & 4.3 $\times 10^{-20}$ & \ce{N2} : \ce{H2O} = 1 : 10 & 12    & $f$ \\ [0.5ex]
\ce{O2}    &                                               & 1.0 $\times 10^{-19}$ & \ce{H2O} : CO : \ce{CO2} : \ce{O2} = 2 : 2 : 0.5 : 1  & 10    & $g$ \\ [0.5ex]
\bottomrule
\end{tabular}
\tablefoot{\tnote{(a)}\citet{Hagen1981}, \tnote{(b)}\citet{Gerakines1995}, \tnote{(c)}\citet{1993ApJS...86..713H}, \tnote{(d)}\citet{D'Hendecourt1986}, \tnote{(e)}\citet{Jiang1975}, \tnote{(f)}\citet{Bernstein1999}, \tnote{(g)}\citet{Ehrenfreund1992}.}
\end{threeparttable}
\end{table*}

\subsection{Preparation of the pre-stellar icy mantle analogues}
\begin{figure}[!htp]
    \centering
    \includegraphics[width=\hsize]{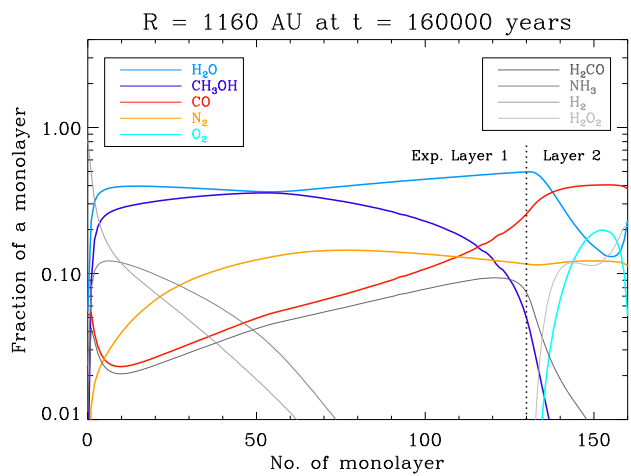}
    \caption{Predicted molecular fractional abundances for each layer of an ice mantle in a region close to the dust peak. 0 on the X-axis corresponds to the grain surface. The vertical line marks the division of monolayers that were used to calculate the average composition in the experimental layers 1 and 2. Coloured lines highlight the species used in our experiments.}
    \label{L1544_dust_peak}
\end{figure}

\indent \indent The pre-stellar icy mantle analogues were  prepared based on the predictions of the gas-grain chemical model of \citet{Vasyunin2017}, updated as explained in Section \ref{model}. The model provides the layer-by-layer molecular composition at different positions within the pre-stellar core.
Figure \ref{L1544_dust_peak} shows an overview of the predicted molecular abundances for each monolayer of the icy mantles present at a radius of 1160~au, a representative region of the central kernel \citep[cf. Fig. 5 in][]{Vasyunin2017}.

In the case of our layered ice analogue, we divided the modelled data into two parts. The molecular composition for the experimental first layer was obtained by averaging over the predicted inner 131 MLs where high fractions of \ce{H2O} and \ce{CH3OH} are present. The three most abundant species (\ce{H2O}, \ce{CH3OH}, and \ce{N2}) were used for the ice mixture of layer 1. For the second layer, we averaged over the outer 30 MLs with high abundances of CO and \ce{O2}. Again, the three most abundant species (CO, \ce{H2O}, and \ce{N2}) as well as \ce{O2} were incorporated into the ice mixture.
Similarly, we averaged over all modelled ice layers and used the four most abundant species (\ce{H2O}, \ce{CH3OH}, CO, and \ce{N2}) for the mixed ice.
A detailed overview of the species present in the model is shown in Table \ref{mod_and_rel_water} where molecular species used for the experiments are highlighted. Table \ref{exps_real} shows the actual ice composition during the experiments, with deviations $\precsim$~10~\% from the predicted ratios.

\begin{table*}[!htp]
\centering
\begin{threeparttable}
\caption{Average fraction of molecules in the ice mantles based on the presented model. The species in bold were used for the ice compositions. The values in brackets give the molecular abundances relative to \ce{H2O}.}
\label{mod_and_rel_water}
\begin{tabular}{c || c c | c}
\toprule \toprule
           & \multicolumn{2}{c}{Layered}                                           & Mixed      \\[1ex]
Molecule   & Layer 1                                       & Layer 2               &            \\
           & (ML 0-130)                                    & (ML 131-160)          & (ML 0-160) \\
\midrule
\ce{H2O}   & $\hskip -0.3em${\bf 40\% (100\%)}             & $\hskip 0.5em${\bf 26\% (100\%)}              & {\bf 38\% (100\%)}                \\
\ce{CH3OH} & {\bf 27\% (67.5\%)}                           & 0.6\% (2.3\%)                                 & $\hskip -0.5em${\bf 22\% (58\%)}  \\ 
\ce{N2}    & {\bf 11\% (27.5\%)}                           & {\bf 12\% (46\%)}                             & $\hskip -0.5em${\bf 11\% (29\%)}  \\ 
\ce{CO}    & $\hskip -0.1em$8\% (20\%)                     & $\hskip 0.5em${\bf 38\% (146\%)}              & $\hskip -0.5em${\bf 14\% (37\%)}  \\ 
\ce{H2CO}  & $\hskip 0.75em$5\% (12.5\%)                                  & $\hskip 0.75em$2\% (7.7\%)                    & 5\% (13\%)         \\ 
\ce{NH3}   & 4\% (10\%)                                                   & 0.1\% (0.4\%)                                 & 3\% (7.9\%)        \\ 
\ce{H2}    & 4\% (10\%)                                                   & 2.5$\times 10^{-4}$\% (9.6$\times 10^{-4}$\%) & 3\% (7.9\%)        \\ 
\ce{H2O2}  & $\hskip -0.5em$1.5$\times 10^{-3}$\% (3.8$\times 10^{-3}$\%) & $\hskip 0.75em$11\% (42.3\%)                  & 2\% (5.3\%)        \\ 
\ce{O2}    & $\hskip -0.5em$2.6$\times 10^{-4}$\% (6.5$\times 10^{-4}$\%) & $\hskip 0.75em${\bf 10\% (38.5\%)}            & 2\% (5.3\%)        \\[0.5ex]
\bottomrule
\end{tabular}
\end{threeparttable}
\end{table*}

\begin{table*}[!htp]
\centering
\begin{threeparttable}
\caption{Ice composition during the experiments compared to predicted ratios based on the model. The measured column density $N$, corresponding ice thickness $d$, and consequential relative ratios have an error of 20~\% -- 30~\% for strong bands (\ce{H2O}, \ce{CH3OH}, and CO) and 50~\% for week bands (\ce{N2} and \ce{O2}).}
\label{exps_real}
\begin{tabular}{l c | c | c c | c c | c c | c c | c c}
\toprule \toprule
        &         & $\chi_{\ce{H2O}}$\tnote{a} & \multicolumn{2}{c}{$\chi_{\ce{CH3OH}}$\tnote{a}} & \multicolumn{2}{c}{$\chi_{\ce{CO}}$\tnote{a}} & \multicolumn{2}{c}{$\chi_{\ce{N2}}$\tnote{a}} & \multicolumn{2}{c}{$\chi_{\ce{O2}}$\tnote{a}} & $N$/10$^{18}$   & $d$   \\[1ex]
        &         &       & Exp.   & Model  & Exp.    & Model & Exp.   & Model  & Exp.   & Model      & (mol cm$^{-2}$) & (MLs) \\
\midrule
Layered & Layer 1 & 100\% & 74\%   & 67.5\% & --      & --    & 32\%   & 27.5\% & --     & --         & 1.08            & 870   \\[0.5ex]
        & Layer 2 & 100\% & --     & --     & 144\%   & 146\% & 48\%   & 46\%   & 41\%   & 38\%       & 0.22            & 180   \\
\midrule
Mixed   &         & 100\% & 57\% & 58\%     & 33\%    & 37\%  & 32\%   & 29\%   & --     & --         & 1.47            & 1180  \\
\bottomrule
\end{tabular}
\begin{tablenotes}
\item [a] Abundance of the molecular species in the ice with respect to \ce{H2O}.
\end{tablenotes}
\end{threeparttable}
\end{table*} 

We note that \ce{CO2} is not a major component of the icy mantle composition predicted by our chemical model. Although this may not be correct, as \ce{CO2} ice is clearly detected in molecular clouds, here we decided  to focus on the exact model predictions relevant for the centre of pre-stellar cores.

\section{Results}
\label{results}
\indent\indent The results for the layered and mixed ices are presented in the next subsection, followed by a more detailed depiction of the weak \ce{N2} and \ce{O2} bands.

\subsection{Layered and mixed ices}
\label{layers}
\indent\indent The experiments were designed in order to fit the molecular composition and the thickness ratio, estimated in number of monolayers, as closely as possible to the model, for the inner and outer ice layers. 
We expect a high error for the weak \ce{N2} and \ce{O2} bands as their band strengths strongly depend on the ice mixture and the relative amounts of polar and apolar components, and we assume an error of 50~\% for the sake of comparison.
A minimum column density of 0.22~$\times$~10$^{18}$~mol~cm$^{-2}$ was necessary to reconfirm a \ce{N2}:\ce{H2O} ratio in layer 2 within this error, while the ML ratio of layer 1 to layer 2 varies less than 10~\% from the modelled partition.

The spectra of layer 1 and layer 2 were recorded in separate experiments. Then, after confirming that the relative abundances in the two ice layers were as close as possible to the modelled predictions, the very same mixtures were used to measure the spectrum of layer 2 on top of layer 1 in a third dedicated experiment.

\begin{figure*}[!htp]
    \centering
    \includegraphics[width=\hsize]{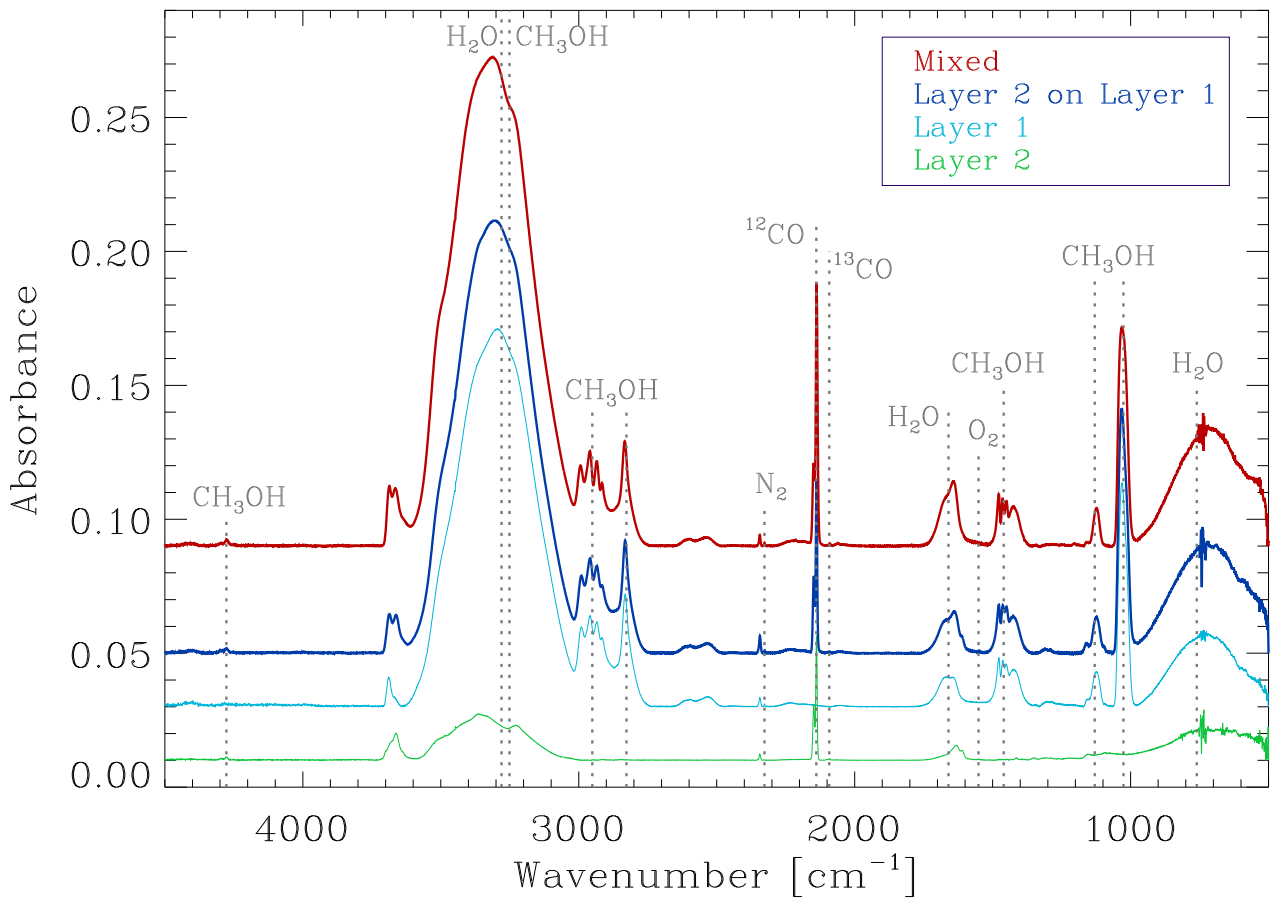}
    \caption{Spectra of mixed ice (red), and layered ices with layer 1 (light blue), layer 2 (green), and layer 2 on top of layer 1 (dark blue). Corresponding to Table \ref{mod_and_rel_water}, the relative abundances for the mixed ice are \ce{H2O}:\ce{CH3OH}:\ce{N2}:CO = 100\%:58\%:29\%:37\% and for the layered ice \ce{H2O}:\ce{CH3OH}:\ce{N2} = 100\%:67.5\%:27.5\% (layer 1) and \ce{H2O}:\ce{N2}:CO:\ce{O2} = 100\%:46\%:146\%:38\% (layer 2).}
    \label{layer_mix_all}
\end{figure*}

Figure \ref{layer_mix_all} shows all the recorded spectra in the whole frequency range for the inner layer 1, the outer layer 2, layer 2 deposited on top of layer 1, and the mixed ice. The positions of band maxima for pure \ce{H2O}, \ce{CH3OH}, and CO ices as well as \ce{N2} and \ce{O2} embedded in a \ce{H2O} matrix are included to help the reader assign the bands to the respective species.
The spectra in all presented figures were baseline corrected using the rubberband correction method offered by the OPUS spectroscopy software, and shifted along the y-axis for a better comparison of mixed and layered ices and to enable a better understanding of how layers 1 and 2 contribute to the spectral features observed in the layered ice.

\subsubsection{\ce{H2O} ice bands}
\label{H2O}
\begin{figure*}[!htp]
    \centering
    \vspace{0.5cm}
    \subfloat{\includegraphics[scale=.5]{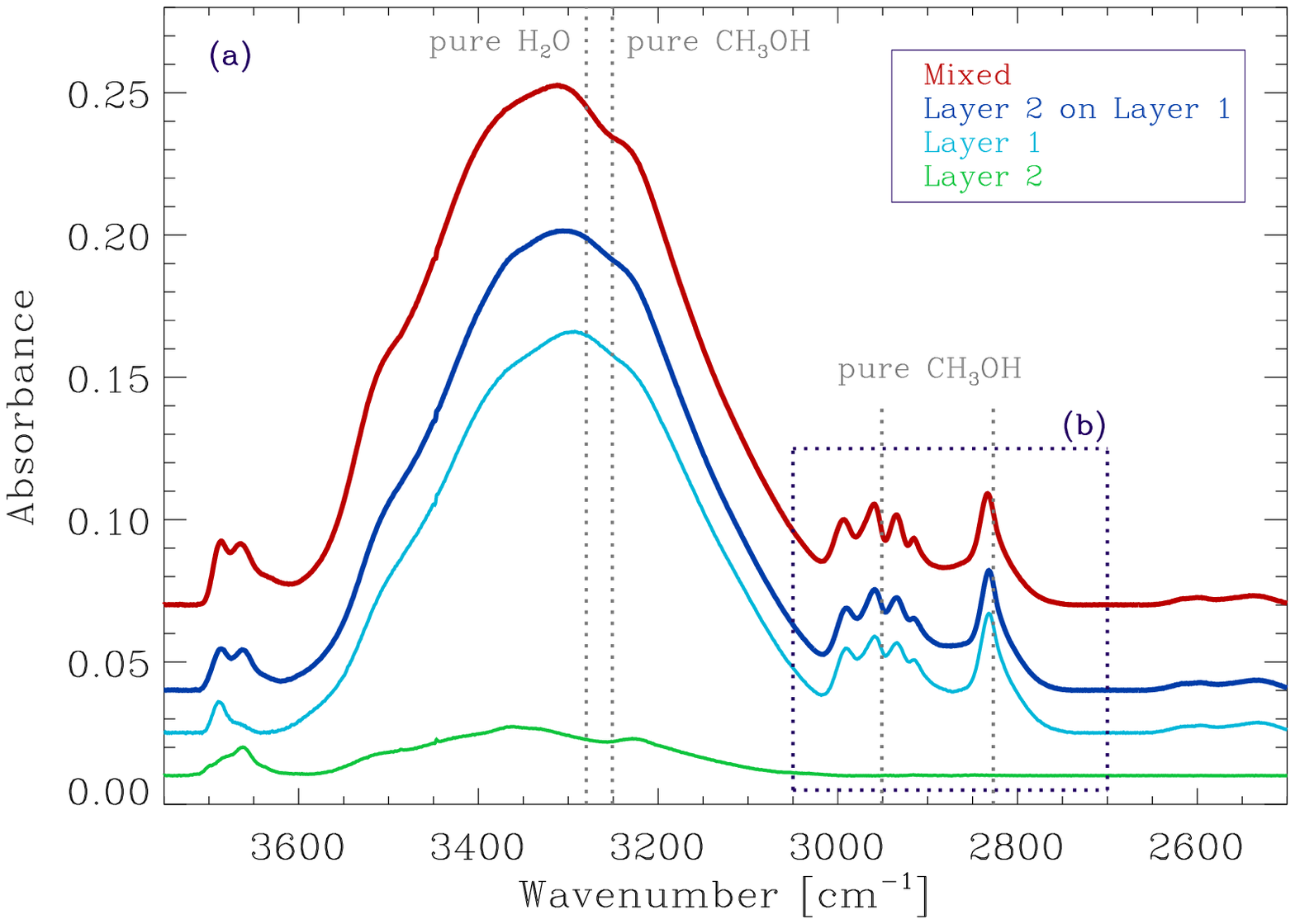}}
    \subfloat{\includegraphics[scale=.5]{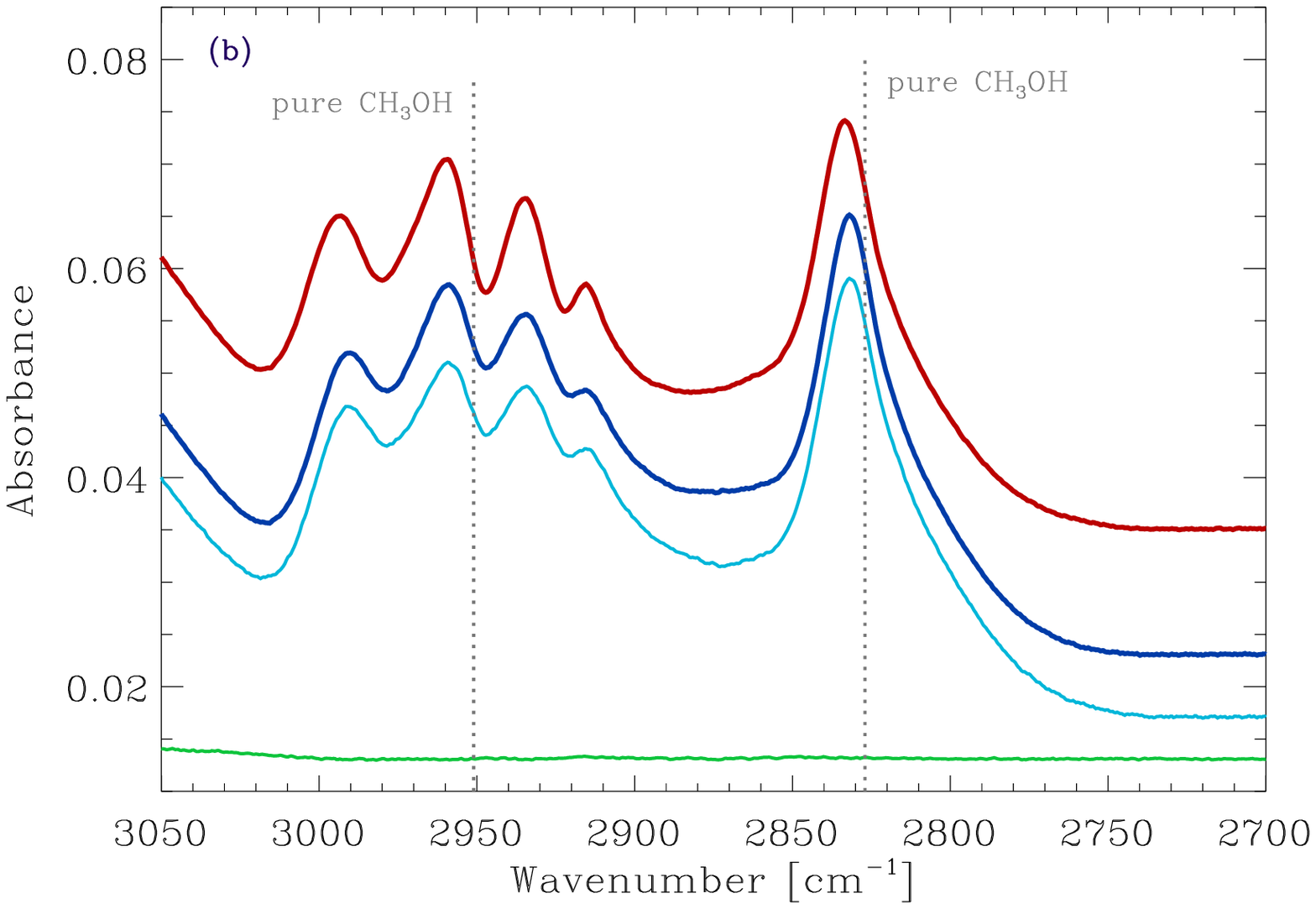}}
    \vspace{0.5cm}
    \subfloat{\includegraphics[scale=.5]{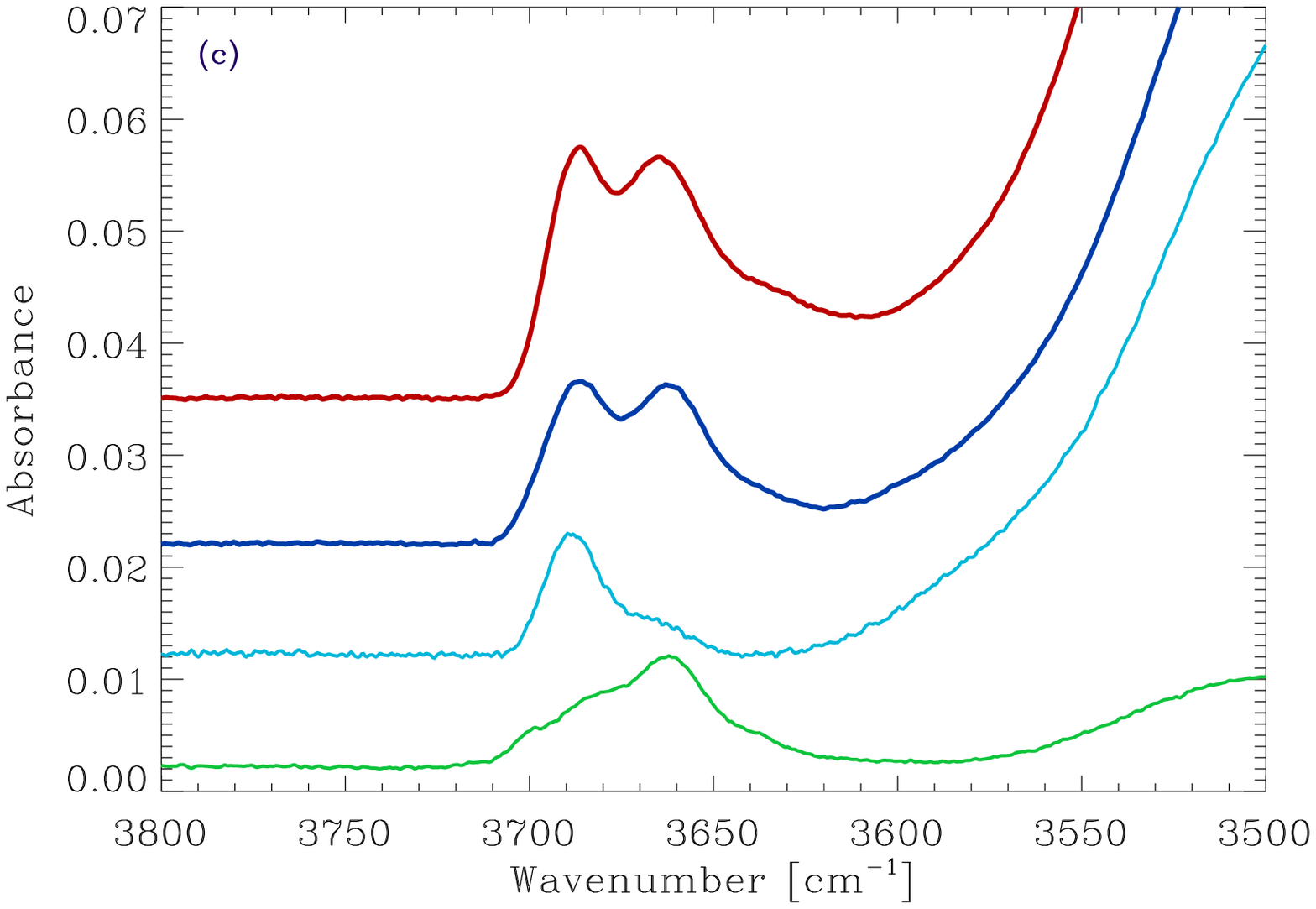}}
    \subfloat{\includegraphics[scale=.5]{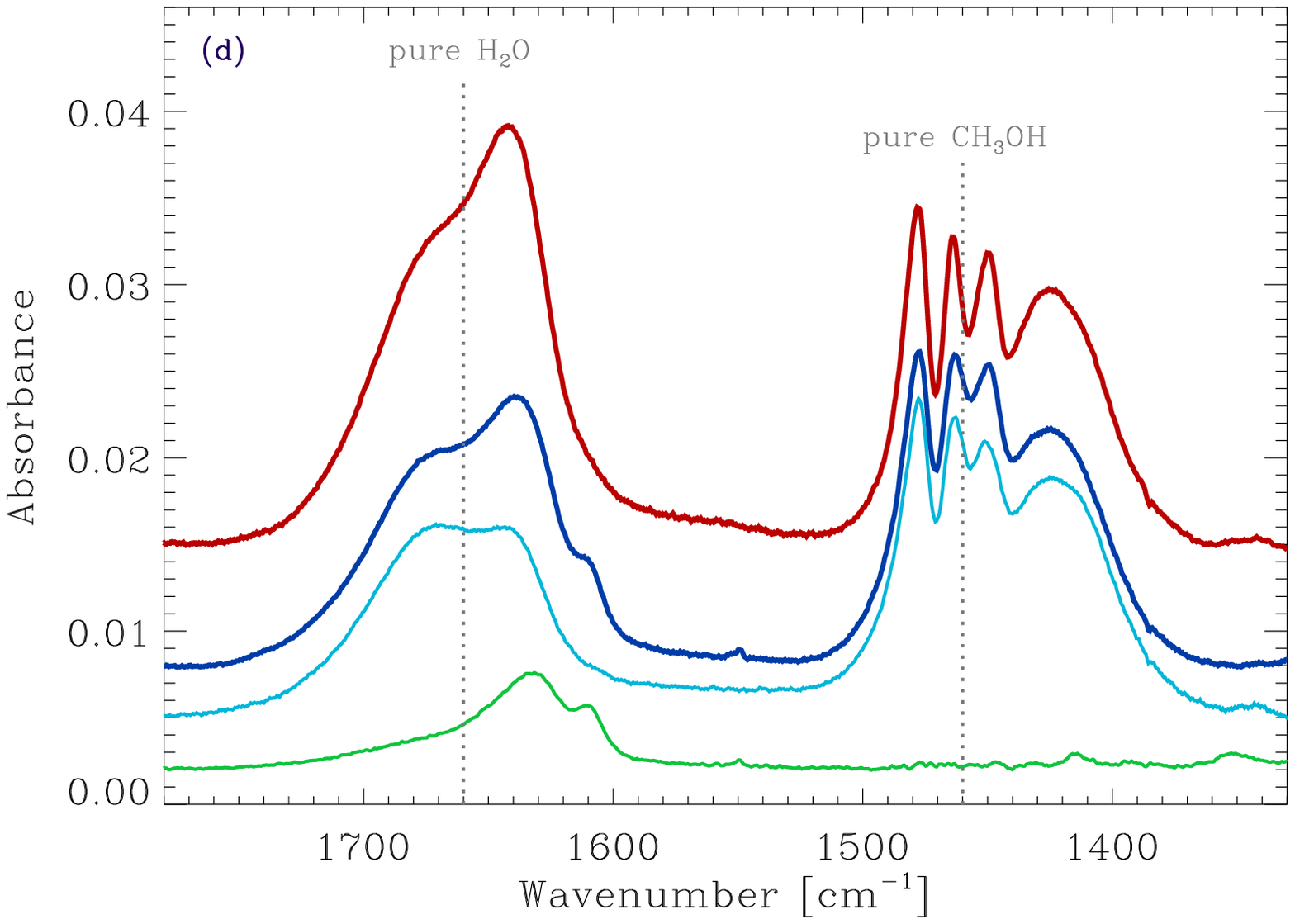}}
    \vspace{0.5cm}
    \subfloat{\includegraphics[scale=.5]{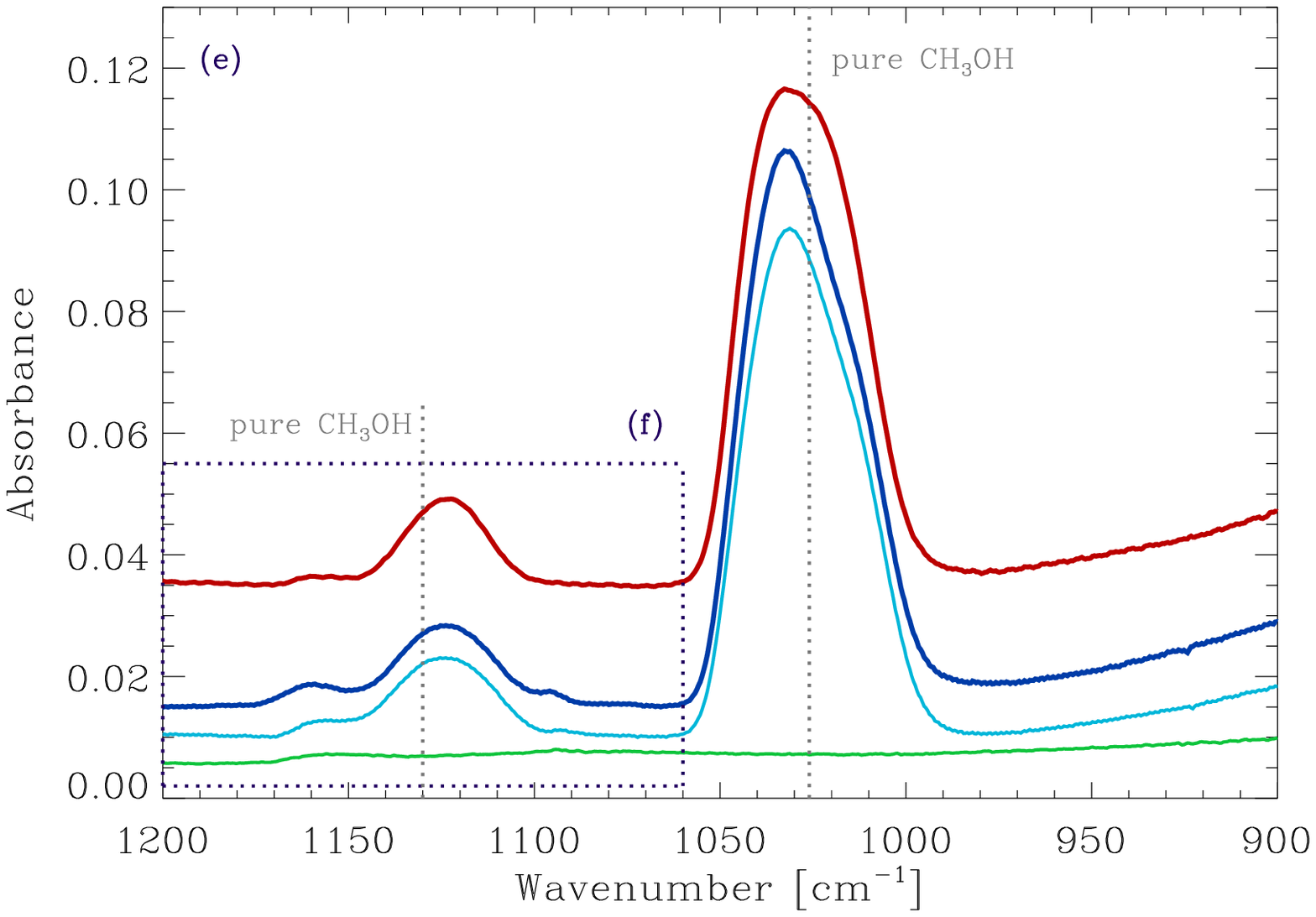}}
    \subfloat{\includegraphics[scale=.5]{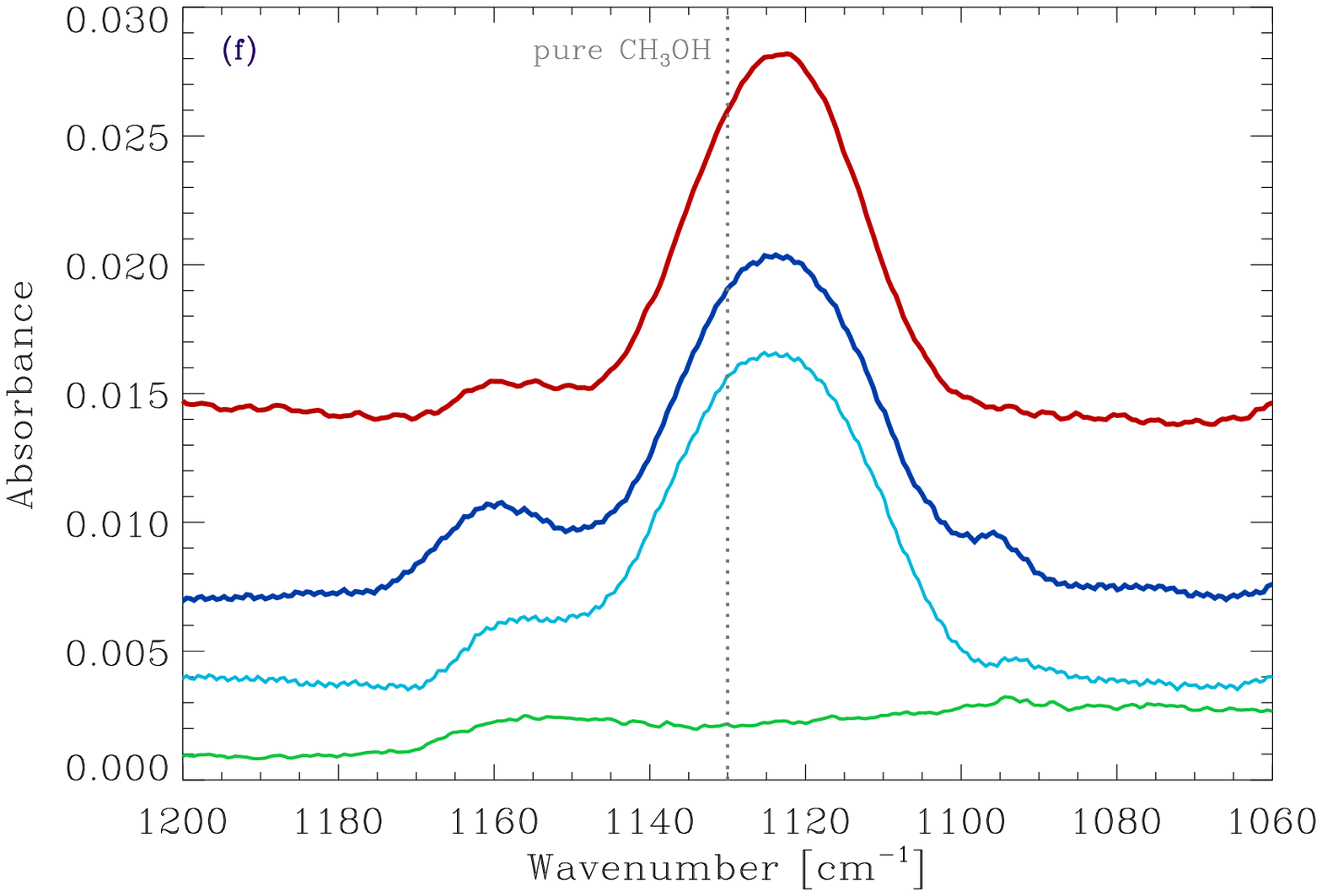}}
    \caption{Spectra of mixed ice (red), and layered ices with layer 1 (light blue), layer 2 (green), and layer 2 on top of layer 1 (dark blue), zoomed-in around the OH-stretch (upper left), CH-stretch (upper right), \ce{H2O} dangling bonds (middle left),  OH-bend (middle right), \ce{CH3OH} CO-stretch (lower left), and \ce{CH3}-rock (lower left) bands. Corresponding to Table \ref{mod_and_rel_water}, the relative abundances for the mixed ice are \ce{H2O}:\ce{CH3OH}:\ce{N2}:CO = 100\%:58\%:29\%:37\% and for the layered ice \ce{H2O}:\ce{CH3OH}:\ce{N2} = 100\%:67.5\%:27.5\% (layer 1) and \ce{H2O}:\ce{N2}:CO:\ce{O2} = 100\%:46\%:146\%:38\% (layer 2).}
    \label{layer_mix_all_h2o_ch3oh}
    \vspace{0.5cm}
\end{figure*}

\indent\indent Figure \ref{layer_mix_all_h2o_ch3oh}a shows a shift in the OH stretching band maximum towards higher frequencies for both mixed and layered ices when compared to the pure ices. While a weak right shoulder feature can be explained by the contribution of the \ce{CH3OH} OH stretching vibration and by interactions with CO, we observe that the  shift in the band maximum is bigger for higher CO:\ce{H2O} and \ce{O2}:\ce{H2O} ratios, as most prominently seen in the spectrum of layer 2.

The molecular composition has a strong effect on the position of the \ce{H2O} dangling bonds as well. The presence of \ce{CH3OH} which has a similar abundance in the mixed ice and layer 2, causes a shift from higher frequencies to 3686~cm$^{-1}$ and an increasing amount to CO leads to dangling bond features even closer to the OH stretching band, as seen in Fig. \ref{layer_mix_all_h2o_ch3oh}c.

In the case of the \ce{H2O} 1660~cm$^{-1}$ bending mode, we see conspicuous differences  in band shape and in position depending on the species that are interacting with \ce{H2O}. Even though the total amount of MLs is similar for the mixed and layered ices, the band intensity as well as the area below the band is considerably smaller for the layered ice.
Additionally, the interaction of \ce{CH3OH} with \ce{H2O} in layer 1 causes a split of the band into two components, in comparison with the spectral feature of the pure \ce{H2O} \citep[cf.][]{Hagen1981}.
On the other hand, in layer 2 with high abundances of CO and \ce{O2}  we observe a shift in the water bending mode towards lower frequencies and the emergence of a shoulder feature on the right wing, as seen in the right panel of Fig. \ref{layer_N2_O2}.

\subsubsection{\ce{CH3OH} ice bands}
\label{CH3OH}
\indent\indent As the average composition of layer 1 was calculated by using 80~\% of the modelled dust grain monolayers, the abundances of \ce{CH3OH} and \ce{N2} relative to \ce{H2O} are very similar to those in the mixed ice.
The major difference between these two ice analogues that  contain  \ce{CH3OH}   is the presence of CO in the mixed ice rather than in layer 1 as it can be found in high amounts only in the outer layers of our model, and thus is incorporated only in the mixed ice and layer 2, where the latter shows a considerably higher CO abundance relative to \ce{H2O}.

Since the \ce{CH3OH}:\ce{H2O} ratio is only slightly higher in layer 1 than in the mixed ice, interactions with CO are presumably the main reasons for the minor shifts in position of the \ce{CH3OH} CH stretching modes in the layered and mixed ices, at 2951~cm$^{-1}$ and 2827~cm$^{-1}$, as seen in Fig. \ref{layer_mix_all_h2o_ch3oh}b.
When compared to pure \ce{CH3OH} ice however \citep[e.g.][]{1993ApJS...86..713H}, we see the effect of the interacting \ce{H2O} and \ce{CH3OH} molecules in both the band shape and the position. The maximum position is shifted to lower frequencies and local maxima appear to be more prominent for the 2951~cm$^{-1}$ band \citep[cf.][]{Ehrenfreund1999}.
In the mixed ice, the presence of CO has a noticeable effect on the intensity, which is enhanced.

A similar behaviour is seen in Fig. \ref{layer_mix_all_h2o_ch3oh}d for the 1460~cm$^{-1}$ \ce{CH3} deformation band. While the band strengths become more prominent with the presence of CO, the general appearance of the band for the mixed ice and layer 1 is very much alike. Interaction of \ce{H2O} and \ce{CH3OH} on the other hand causes not only a small blueshift, but also a change in the band shape. A comparison with Fig. 13 in \citet{Ehrenfreund1999} shows that the relative intensity of the three local maxima at higher frequencies ($\nu_4$, $\nu_5$, $\nu_6$) changes, with the left wing becoming the most intense. The shoulder feature on the right wing of the 1460~cm$^{-1}$ methanol band ($\nu_{10}$) also becomes more dominant.

In Figs. \ref{layer_mix_all_h2o_ch3oh}e and \ref{layer_mix_all_h2o_ch3oh}f no shifts in position of the CO stretching and \ce{CH3} rocking bands are seen when comparing layered and mixed ices. While a small change in the 1026~cm$^{-1}$ band shape might be difficult to observe, the emergence of a local maximum on the right wing of the 1130~cm$^{-1}$ band in the layered ice, as well as the different shape of the left wing local maximum, enable differentiation of the layered or mixed structure of the ice.
Additionally, the interaction of \ce{H2O} and \ce{CH3OH} leads to a shift towards higher frequencies relative to the position of pure \ce{CH3OH} ice for the CO stretching mode, while the \ce{CH3} rock shifts towards smaller wavenumbers.

\subsubsection{CO ice band}
\label{CO}
\indent\indent The presented model suggests that a considerable amount of CO freezes out within pre-stellar cores, in agreement with observations \citep[e.g.][]{Crapsi2005}. CO is the most abundant species in the external layers of icy mantles in contrast with the composition of the bulk in which \ce{H2O} represents the largest fraction. In our experiments this layered morphology of CO on top of \ce{H2O} is represented in layers 1 and 2 and is compared to the mixed ice with a considerably lower CO:\ce{H2O} ratio. 

Figure \ref{layer_mix_all} shows that position and shape of the CO ice bands are very similar in the layered and mixed ice analogues. For both spectra we detect only a minor shift from 2139~cm$^{-1}$ to 2138~cm$^{-1}$ when compared to pure CO ice \citep[e.g.][]{Jiang1975}. A second component is visible in the absorption band, due to the interaction with \ce{H2O}, for layered and mixed ices at 2148~cm$^{-1}$ or 2149~cm$^{-1}$, respectively \citep[cf.][]{Bouwman2007}. Band shape and position are not affected by the presence of \ce{CH3OH}, \ce{N2}, and \ce{O2}.
Thus, we find that CO spectral features do not bring any information about the layered or mixed nature of our ice analogues. 

\subsection{\ce{N2} and \ce{O2}}
\label{diatomic}
\indent\indent Pure \ce{N2} and \ce{O2} do not show vibrational bands in the IR region because homonuclear diatomic species have no dipole moment. A molecular stretching mode can only be induced via symmetry breaking by other polar species such as water, as shown by \citet{Ehrenfreund1992} and \citet{Bernstein1999}.

Molecular nitrogen \ce{N2} is present in all layers of the considered ice mantle. Apart from the innermost bulk, it shows a constant abundance throughout all numbers of monolayers, and thus the average \ce{N2} fractions of both layered and mixed ices are very similar. Nonetheless, in our layered ice experiment the \ce{N2}:\ce{H2O} ratio is higher for layer 2 than for layer 1, which is due to a decreasing abundance of \ce{H2O} towards the surface layers.

\begin{figure*}[!htp]
    \centering
    \vspace{0.5cm}
    \subfloat{\includegraphics[scale=.5]{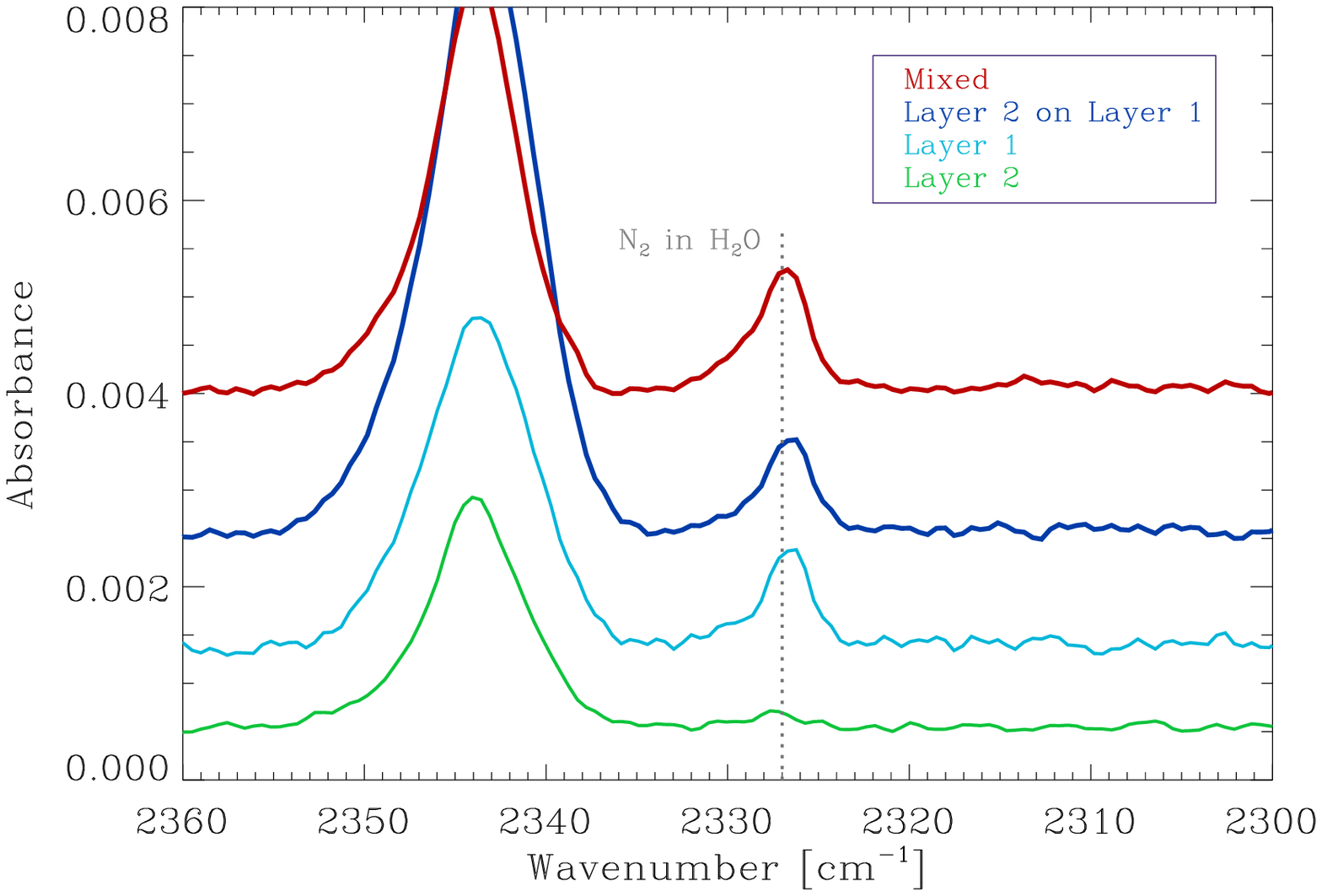}}
    \subfloat{\includegraphics[scale=.5]{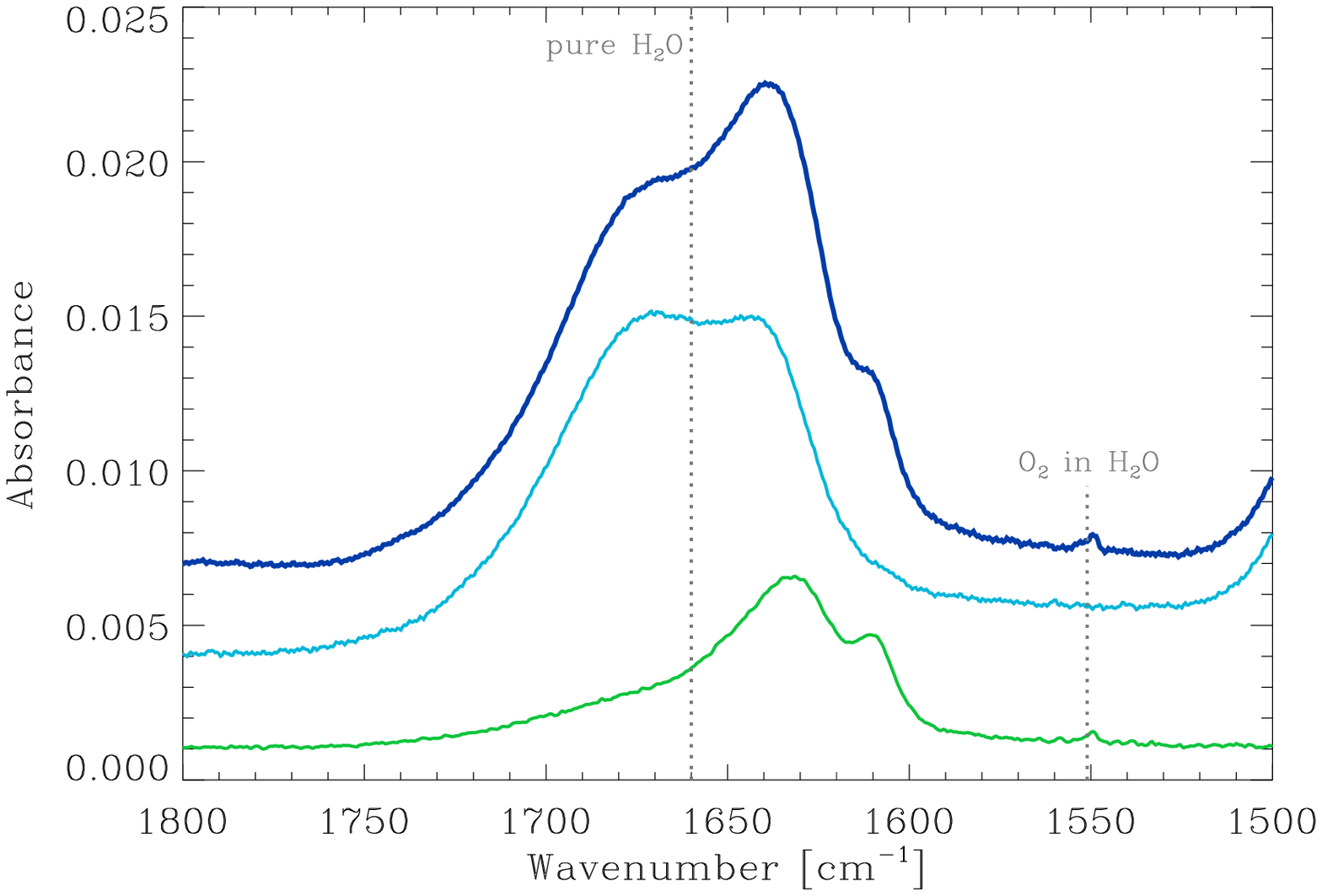}}
    \caption{Spectra of mixed ice (red), and layered ices with layer 1 (light blue), layer 2 (green), and layer 2 on top of layer 1 (dark blue), zoomed-in around the \ce{N2} (left) and \ce{O2} (right) bands. Corresponding to Table \ref{mod_and_rel_water}, the relative abundances for the mixed ice are \ce{H2O}:\ce{CH3OH}:\ce{N2}:CO = 100\%:58\%:29\%:37\% and for the layered ice \ce{H2O}:\ce{CH3OH}:\ce{N2} = 100\%:67.5\%:27.5\% (layer 1) and \ce{H2O}:\ce{N2}:CO:\ce{O2} = 100\%:46\%:146\%:38\% (layer 2).}
    \label{layer_N2_O2}
\end{figure*}

For the relative abundances used in this study, the left panel of Fig. \ref{layer_N2_O2} shows that the differences in the relative abundances of \ce{N2} and \ce{H2O} have no influence on the 2327~cm$^{-1}$ nitrogen band position and shape. The same is true for different ice compositions including \ce{CH3OH}, CO, and \ce{O2} in different composition ratios.

As already mentioned, we did not include \ce{CO2} in our modelled experiments as we followed the theoretical predictions of our pre-stellar core chemical model. Still, a weak \ce{CO2} band is present in Fig. \ref{layer_N2_O2} due to residual contamination in the experimental set-up. 

In the inner regions of L1544, \ce{O2} is predicted to be present in high amounts of 10~\% in the outer monolayers of the ice mantles. With an average abundance of \ce{H2O} = 26~\% in these layers, and a relative abundance of \ce{O2}:\ce{H2O} = 38~\%, the \ce{O2} band becomes visible in layer 2 of our experiments.
We find the \ce{O2} band at a frequency of 1549~cm$^{-1}$, which differs from   1551~cm$^{-1}$ only when \ce{O2} is present   in a two-component \ce{O2}-\ce{H2O} ice mixture \citep[cf.][]{Ehrenfreund1992,Mueller2018}.

\section{Discussion}
\label{discuss}

\subsection{Comparison with previous work}
\indent\indent We have seen that the structure and the molecular composition of the ice have an effect on the position and shape of absorption bands. Table \ref{bands} lists all the examined band positions compared with the position values of pure \ce{H2O}, \ce{CH3OH}, and CO ices, as well as \ce{N2} and \ce{O2} embedded in a water matrix.
Some of the observed bands do not show any shifts when layered and mixed ices are compared. Nevertheless, our results show important deviations from the frequencies measured for pure ices.
In this context, the observed broadening of the 1032~cm$^{-1}$ maximum in the mixed ice indicates the interaction of \ce{CH3OH} with CO, which was studied by \citet{Penteado_2015} who recorded a shift towards higher frequencies when \ce{CH3OH} is interacting with CO. Our measurements show that the band profile of the 1464~cm$^{-1}$ \ce{CH3} deformation mode is another potential indicator of the \ce{CH3OH}-CO interplay. Moreover, the particular profile of the CO double peak appears when CO and \ce{H2O} are present in an ice mixture of low temperature, as observed by \citet{Fraseretal2004}.

\begin{table*}[!htp]
\centering
\begin{threeparttable}
\caption{Position of all observed molecular bands in the experiments presented, recorded at 10~K, with a spectral resolution of 1~cm$^{-1}$. Multiple band assignments refer to several local maxima or shoulder features for one band.}
\label{bands}
\begin{tabular}{l | c | c c c}
\toprule \toprule
               &                                       & \multicolumn{3}{c}{Wavenumber (cm$^{-1}$)} \\[1ex]
Molecule   & Vibration mode                    & Pure             & Layered$^{a}$    & Mixed            \\ [0.5ex]
\midrule
\ce{H2O}   & $\nu_1$, $\nu_3$ OH stretch       & 3280             & 3304             & 3314             \\ [0.5ex]
           & $\nu_2$ OH bend                       & 1660             & 1670, 1638, 1610 & 1674, 1642       \\ [0.5ex]
           & dangling bonds                    & 3720, 3697       & 3686, 3662       & 3686, 3665       \\ [0.5ex]
\ce{CH3OH} & OH \& CH stretch                      & 3251, 2951, 2827 & 3304, 2959, 2832 & 3314, 2960, 2834 \\ [0.5ex]
               & \ce{CH3} deformation              & 1460             & 1464             & 1464             \\ [0.5ex]
               & $\nu_7$, $\nu_{11}$ \ce{CH3} rock & 1130                     & 1160, 1124, 1096 & 1160, 1124       \\ [0.5ex]
               & $\nu_8$ CO stretch                        & 1026             & 1032             & 1032             \\ [0.5ex]
CO             & CO stretch                            & 2139             & 2138, 2148       & 2138, 2149       \\ [0.5ex]
\ce{N2}    &                                               & 2327$^{b}$       & 2327             & 2327             \\ [0.5ex]
\ce{O2}    &                                               & 1551$^{b}$       & 1549             & -                \\ [0.5ex]
\bottomrule
\end{tabular}
\begin{list}{}
\item (a) Layer 2 on top of layer 1.
\item (b) Embedded in a \ce{H2O} matrix.
\end{list}
\end{threeparttable}
\end{table*}

Most of the \ce{H2O} and \ce{CH3OH} bands can be used to characterise the layered or mixed nature of our ices. We are able to connect the band changes to the structure of our probes, in agreement with previous works in the literature \citep[e.g.][]{Wolff2007,Bahr_2008,Oeberg2009}.

Based on the experimental results of \citet{Dawes2016}, who examined band changes of \ce{CH3OH} in different ratios relative to \ce{H2O}, we expected to find the peak of the OH stretching band in a  frequency range between the maximum position of 3280~cm$^{-1}$ for pure water and 3251~cm$^{-1}$ for pure methanol. However, our results  show a blueshift towards frequencies $\geq$ 3304~cm$^{-1}$ and the emergence of a right wing shoulder feature. The appearance of the band can be explained by interactions of CO with the OH mode of \ce{H2O} and \ce{CH3OH}. Our observations agree well with \citet{Bouwman2007} who saw a similar behaviour for ices with an increasing CO:\ce{H2O} ratio, and with experiments conducted by \citet{Mueller2021} where CO is present in ice mixtures containing water and methanol.

In addition,  the appearance of a shoulder feature on the right wing of the \ce{H2O} 1660~cm$^{-1}$ bending mode and the shift in the water dangling bonds towards lower frequencies is in good agreement with observations for similar CO:\ce{H2O} and \ce{O2}:\ce{H2O} ratios conducted by \citet{Bouwman2007} and \citet{Mueller2018}, who examined the interaction of \ce{H2O} with CO and \ce{O2}, respectively. Spectra presented in \citet{Mueller2021} do not show much of an  effect of \ce{CH3OH} on the free OH dangling bonds, but we note that the \ce{CH3OH}:\ce{H2O} ratio in that work is considerably lower than in the presented experiments. Considering reflection absorption infrared spectra (RAIRS) results from \citet{He2018c}, who showed the shift in dangling bonds for a \ce{N2}:\ce{H2O} ratio similar to our mixtures, we deduce the remarkable band profile difference of the local maximum at 3686~cm$^{-1}$ from interactions of water with methanol.

The highly varying amounts of CO is the major difference in mixed and layered ices containing \ce{CH3OH}. Thus, differences in band shape and position of \ce{CH3OH} are mainly due to its interaction with CO. 
While the 2951~cm$^{-1}$ asymmetric CH stretching and 1460~cm$^{-1}$ \ce{CH3} bending mode seem to be enhanced by this interaction, we observe only a significant change in the 1130~cm$^{-1}$ \ce{CH3} rocking mode.
\cite{Mueller2021} found that shoulder features of the 1130~cm$^{-1}$ band tend to disappear when \ce{CH3OH} interacts with increasing abundances of CO. This behaviour can also be seen when we compare our layered ice, where CO and \ce{CH3OH} are present in different layers, with the mixed ice, where the presence of CO and methanol causes the decrease in or even the disappearance of the left and right shoulder features.
Apart from the shift towards lower frequencies due to interactions of \ce{H2O} and \ce{CH3OH}, we do not see any differences in the 2827~cm$^{-1}$ symmetric CH stretch when comparing layered and mixed ices.
The shift in the 1026~cm$^{-1}$ CO stretching band towards higher frequencies agrees well with the results in \cite{Mueller2021}, which reports    a CO:\ce{CH3OH} ratio appropriate for comparison with the present work. However, the present study  cannot confirm the shift in the CO stretching mode to 1018~cm$^{-1}$, as seen in \citet{Mueller2021} for \ce{CH3OH}-\ce{H2O} mixtures when CO is not included in the ice matrix. The discrepancy  can be explained by the large difference in the ratio of \ce{CH3OH}:\ce{H2O} = 5~\% in the previous work, compared to the present work with an ice fraction of \ce{CH3OH}:\ce{H2O} = 74~\%.

Examining the CO stretching band we see that it is not affected by changing \ce{N2}:CO ratios, or the presence of \ce{O2} or \ce{CH3OH} in the mixed ice. However, when we compare our data to  the measurements conducted by \citet{Bouwman2007}, we see that a mixture of CO and \ce{H2O} results in a feature on the left wing of the band, which increases for increasing \ce{H2O}:CO ratios, and shifts towards higher frequencies. Moreover, our observations agree well with the broadening of the band and decreasing intensity as recorded by \citet{Bouwman2007} for increasing abundances of \ce{H2O}.
Comparisons with studies from \citet{Cuppen2011} and \citet{Penteado_2015}, on the other hand, show that the left wing is missing in the spectra of young stellar objects, which  indicates that \ce{H2O} and CO are not intermixed in the observed ices.

\subsection{Ice feature observability with JWST}
\label{JWST_ETC_section}
\indent\indent We are interested in the observability of the weak \ce{N2} and \ce{O2} bands with JWST as both species are thought to be abundant in interstellar ices. Although  not detectable in the IR region in their pure form, both species become visible at mid-IR wavelengths when they are embedded in a polar matrix (for example \ce{H2O}) that can induce the formation of a dipole moment.

The presented model predicts large amounts of molecular nitrogen in all ice layers forming on dust grains in the inner regions of L1544. In our analysis the \ce{N2} feature is not sensitive to the changes in ice composition or structure. This is confirmed by the work of \citet{Bernstein1999} who analysed the nitrogen absorption band in binary mixtures with excess of \ce{N2}.

Molecular oxygen \ce{O2}  is thought to be found in fractions of a monolayer >~10~\% in the outer layers of ice mantles located in central zones of the pre-stellar core. In our experiment, this correlates to a high \ce{O2}:\ce{H2O} ratio of 38.5~\% coming from the modelling results and the band becomes detectable for the FTIR spectrometer. We find the \ce{O2} band at a position of 1549~cm$^{-1}$, which differs from the frequency of 1551~cm$^{-1}$ that was recorded for binary \ce{O2}-\ce{H2O} mixtures by \citet{Ehrenfreund1992} and \citet{Mueller2018} with spectral resolutions of 4~cm$^{-1}$ and 2~cm$^{-1}$, respectively. The difference in position can be explained with the interaction of \ce{O2} with CO \citep[cf.][]{Ehrenfreund1992}

In the case of our layered ice analogue, for some absorption bands we noticed differences in the band shape when we compared layers 1 and 2 with layer 2 deposited on top of layer 1. The differences are present in the water dangling bonds and bending mode and in the methanol \ce{CH3} rocking mode. A comparison of the computationally added layers 1 and 2 with the experimentally layered ice is shown in Appendix  \ref{compar_l_meas_comp}. The differences in band shape and position can be explained by interactions of molecules at the interface between the two layers.

For both the layered and the mixed ice, we estimate the observation times with JWST for a selection of the presented absorption bands, including the weak \ce{N2} and \ce{O2} bands. The optical depth $\tau$ and the transmittance $T$ are related to each other by
\begin{equation}
 T = \mathrm{e}^{- \tau} = \mathrm{e}^{- \int \alpha~ \mathrm{d}s},
\end{equation}
where d$s$ is the length of the path through which the light of a background star travels. Similarly to what was found in  \citet{Mueller2018}, the absorption coefficient $\alpha$ can be derived from absorbance values $Abs$ and ice thicknesses $d$ obtained during the experiments using

\begin{equation}
\alpha = \frac{Abs \cdot \mathrm{ln}(10)}{d}.
\end{equation}

For our calculations of the whole pre-stellar core we sum up the optical depth of subpaths $s_n$ along the line of sight and derive
\begin{equation}
 \tau_{\mathrm{tot}} =  \sum_{\mathrm{s_n}} \tau_{\mathrm{s_n}}
,\end{equation}
and consequently
\begin{equation}
 T_{\mathrm{tot}} = \prod_{\mathrm{s_n}} T_{\mathrm{s_n}}.
\end{equation}In order to derive the total optical depth and transmittance of the modelled pre-stellar core, the ice composition and thickness at each subpath $s_n$ is taken into consideration. The model suggests that the proportions of icy species within a radius of 1675~au do not change much and that this is also true for ice mantles in the outer region of the modelled L1544. Thus, we decided to treat the ice mantles in the inner 1675~au as the same as our experimentally tested ice composition, while we average over all chemical fractions for R > 1675~au. The calculated values for $\tau$ and $T$ are shown in Table \ref{tau_Tr}.

\begin{table*}[!htp]
\centering
\begin{threeparttable}
\caption{Optical depth and transmittance of the modelled ice features in L1544. The values have an error of 20~\% -- 30~\% for strong bands (\ce{H2O} and \ce{CH3OH}) and 50~\% for week bands (\ce{N2} and \ce{O2}).}
\label{tau_Tr}
\begin{tabular}{l | c | c c | c c}
\toprule \toprule
               &                                       & \multicolumn{2}{c}{$\tau$} & \multicolumn{2}{c}{$T$} \\[1ex]
Molecule   & Vibration mode                    & Layered & Mixed            & Layered & Mixed   \\ [0.5ex]
\midrule
\ce{H2O}   & $\nu_1$, $\nu_3$ OH stretch       & 3.81  & 3.82               & 0.022  & 0.022  \\ [0.5ex]
           & $\nu_2$ OH bend                       & 0.37  & 0.51               & 0.69  & 0.60  \\ [0.5ex]
\ce{CH3OH} & OH \& CH stretch                      & 3.81  & 3.82               & 0.022  & 0.022  \\ [0.5ex]
               & $\nu_7$, $\nu_{11}$ \ce{CH3} rock & 0.32  & 0.30               & 0.73  & 0.74  \\ [0.5ex]
               & $\nu_8$ CO stretch                        & 2.16  & 1.70               & 0.12  & 0.18  \\ [0.5ex]
\ce{N2}    &                                               & 0.022 & 0.026              & 0.98  & 0.97  \\ [0.5ex]
\ce{O2}    &                                   & 0.018 & -                  & 0.98  & -       \\ [0.5ex]
\hline
\end{tabular}
\end{threeparttable}
\end{table*}

The transmitted fraction of the initial flux density $F_{\nu}^{\star}$ emitted by a background star is
\begin{equation}
F_{\mathrm{\nu,tr}} = T \cdot F_{\nu}^{\star}
\end{equation}
and the absorbed fraction is thus 
\begin{equation}
F_{\mathrm{\nu,abs}} = A \cdot F_{\nu}^{\star} = (1-T) \cdot F_{\nu}^{\star}.
\end{equation}Finally, the line strength necessary for calculating the observation time is
\begin{equation}
S = \int_{\Delta\nu} F_{\mathrm{\nu,abs}}~d\nu,
\end{equation}
where $\Delta\nu$ is the band width of the corresponding absorption band. Assuming that the band has the form of a Gaussian peak, the line strength is approximately $S \approx F_{\mathrm{\nu,abs}} \cdot FWHM$, where $FWHM$ is the full width at half maximum.

In order to calculate the line strengths close to the dust peak, which are necessary to obtain observation times with JWST, the apparent magnitudes from observations of L1544 with the Spitzer Space Telescope (SST) were used to derive the flux densities for the SST Infrared Array Camera (IRAC) channels 2-4, which are suitable for observing the 1600~cm$^{-1}$ \ce{H2O} bend, 1130~cm$^{-1}$ \ce{CH3OH} rock, 2327~cm$^{-1}$ \ce{N2}, and 1551~cm$^{-1}$ \ce{O2} bands. Close to the dust peak the observed Vega magnitudes are m$_{\mathrm{vega,IRAC2}}$ = 10.61, $m_{\mathrm{vega,IRAC3}}$ = 10.24, and $m_{\mathrm{vega,IRAC4}}$ = 10.16. Converting to AB magnitudes, which are necessary for calculating the flux densities, gives $m_{\mathrm{AB,IRAC2}}$ = $m_{\mathrm{vega,IRAC2}}$ + 3.26 = 13.87, $m_{\mathrm{AB,IRAC3}}$ = $m_{\mathrm{vega,IRAC3}}$ + 3.73 = 13.97, and $m_{\mathrm{AB,IRAC4}}$ = $m_{\mathrm{vega,IRAC4}}$ + 4.40 = 14.56. The resulting flux density consequently is
\begin{equation}
F_{\nu}^{\star} = 10^{-m_{\mathrm{AB}}/2.5} \cdot F_{\nu0},
\end{equation}
where $F_{\nu0}$ is the zero magnitude flux density, which   in the case of SST is equal to $F_{\mathrm{\nu0,IRAC2}}$ = 179.7~Jy, $F_{\mathrm{\nu0,IRAC3}}$ = 115.0~Jy, and $F_{\mathrm{\nu0,IRAC4}}$ = 64.9~Jy.

Together with the $FWHM$ of the bands, the resulting line strengths can be found in Table \ref{obs_time}.
Both quantities are necessary to generate a theoretical spectrum of the modelled L1544 with the bands listed in Table \ref{tau_Tr} and to calculate the observation times of the absorption bands using the JWST exposure time calculator (ETC). The observation times were calculated for signal-to-noise ($S/N$) ratios > 10 for the weak \ce{N2} and \ce{O2} bands and are also in  Table \ref{obs_time}. We find that the detection of \ce{N2} is feasible with a calculated total exposure time of 1.4~h with the Near-Infrared
spectrograph (NIRSpec), while for observing \ce{O2} an exposure time 123.6~h is necessary.
As the \ce{H2O} bending mode is observed with the same instrument configurations as \ce{O2}, its observation time amounts to the same time interval even though it can be detected after shorter exposure times. However, this specific feature can give us information about the layered or mixed nature of icy mantles and  about abundances of CO and \ce{O2} relative to \ce{H2O}. Thus, longer observation times improve the evaluation of the physical and chemical structure of interstellar ices.

\begin{table*}[!htp]
\centering
\begin{threeparttable}
\caption{Full width half maximum $FWHM$, line strength $S$, observation time $t_{\rm{obs}}$, and signal-to-noise ratio $S/N$ of the bands included in the JWST exposure time calculator.}
\label{obs_time}
\begin{tabular}{l | c | c c | c c | c | c c}
\hline \hline
               &                                       & \multicolumn{2}{c}{$FWHM$ (Hz)}               & \multicolumn{2}{c}{$S$ (erg s$^{-1}$~cm$^{-2}$)} & $t_{\mathrm{obs}}$ (s) & \multicolumn{2}{c}{$S/N$} \\[1ex]
Molecule   & Vibration mode                    & Layered              & Mixed                & Layered               & Mixed                  &              & Layered & Mixed \\ [0.5ex]
\midrule
\ce{H2O}   & $\nu_2$ OH bend                   & 2.28$\times 10^{12}$ & 2.04$\times 10^{12}$ & 2.08$\times 10^{-15}$ & 2.40$\times 10^{-15}$  & 445105$^{a}$ & 151.47  & 183.48 \\ [0.5ex]
\ce{CH3OH} & $\nu_7$, $\nu_{11}$ \ce{CH3} rock & 9.00$\times 10^{11}$ & 1.05$\times 10^{12}$ & 2.38$\times 10^{-16}$ & 2.63$\times 10^{-16}$  & 445105$^{b}$ & 41.92   & 40.05 \\ [0.5ex]
\ce{N2}    &                                   & 9.00$\times 10^{10}$ & 9.00$\times 10^{10}$ & 9.87$\times 10^{-18}$ & 1.17$\times 10^{-17}$  & 5017$^{c}$   & 11.23   & 12.72 \\ [0.5ex]
\ce{O2}    &                                   & 9.00$\times 10^{10}$ & -                    & 4.72$\times 10^{-18}$ & -                      & 445105$^{a}$ & 10.21   & - \\ [0.5ex]
\hline
\end{tabular}
\begin{tablenotes}{}
\item [a] Instrument configuration: MIRI MRS Channel 1B.
\item [b] Instrument configuration: MIRI MRS Channel 2B.
\item [c] Instrument configuration: NIRSpec fixed slit G395H/F290LP S400A1.
\end{tablenotes}
\end{threeparttable}
\end{table*}

\balance

The observation time of \ce{O2} is too long to be performed with JWST. However, this might change when a background light source closer to the dust peak and thus more suitable for \ce{O2} detection is used for the calculation.
Future surveys could help to find such a source and make more promising predictions on the detectability of absorption bands.

The purpose of the  results we present here  is providing spectroscopic data that will help future observations to investigate the nature of interstellar ice structures. The available spectral resolutions of $R$ > 2400 for the Mid-Infrared Instrument (MIRI) and $R$ $\sim$ 2700 for NIRSpec (high resolution gratings) correspond to an upper limit of 0.43~cm$^{-1}$ < $\Delta\nu$ < 0.83~cm$^{-1}$ for the examined bands observed with MIRI and 0.79~cm$^{-1}$ < $\Delta\nu$ < 1.38~cm$^{-1}$ for bands observed with NIRSpec. Hence, all changes in the spectroscopic features recorded in our experiments will be detectable for JWST.

Observations with JWST will not only be able to test advanced gas-grain model predictions, but will also constrain the physical and chemical structure of ice mantles. The \ce{H2O} stretching, bending, and dangling; the  \ce{CH3OH} rocking;  and the  \ce{O2} modes help to differentiate between layered and mixed ice structures, whereas composition ratios can be estimated via the \ce{CH3OH} CH stretching and \ce{CH3} bending modes. We conclude that JWST observations will be sufficient to test our model predictions.

\section{Conclusion}
\label{concl}

\indent\indent With this work we wanted to provide a first attempt to execute a three-dimensional fit between the observations, models, and laboratory work instead of the two-dimensional fits that  various works in the literature show. In order to realise this approach we fitted observational results with the gas-grain chemical model predicting the exact ice composition including IR inactive species. Then the laboratory IR spectra were recorded for the interstellar ice analogues whose composition reflects the obtained numerical results. These results then could be compared with future JWST observations.
We paid special attention to the inclusion of IR inactive species whose presence are predicted in the ice, but are systematically omitted in other laboratory obtained data. This first of all is true  for \ce{N2}, one of the main possible constituents of the interstellar ice mantles, and to the some extent \ce{O2}, which is now drawing special attention after detection in the coma of comet 67P.

This work focuses on the laboratory spectroscopy of interstellar icy mantle analogues predicted within pre-stellar cores, and we characterise  the changes in absorption band positions and profiles for ice analogues based on the model of \citet{Vasyunin2017}. Previous \ce{CH3OH} observations by \citet{Dartois_1999}, \citet{Penteado_2015}, and \citet{Goto_2020} hint at a layered structure of ice mantles, as demonstrated in \citet{Mueller2021}.
This work investigates the spectral feature contributions for new species, in addition to \ce{CH3OH}, which are suitable for deducing the morphology of interstellar ices.
The ice structure is closely related to the chemical composition of the mixtures that we prepared for our experiments.
According to our theoretical predictions we find that in regions close to the dust peak, ice layers rich in \ce{H2O} and \ce{CH3OH} are covered by watery layers with a significant amount of CO and \ce{O2}. We compare the IR spectrum of this structure with a uniformly mixed ice and show the differences in the band characteristics.
Our experiments, coupled with future JWST observations, will allow us to test our gas-grain models and help us to gain understanding of surface processes.
The methodology presented in this study has the potential to be applied to future works with astrochemical models which present even closer fits to the observational results.

\begin{acknowledgements}
We thank Christian Deysenroth for the designing and development of the experimental set-up and the continuous assistance in the laboratory development. The work of Birgitta M\"uller was supported by the IMPRS on Astrophysics at the Ludwig-Maximilians University, as well as by the funding the IMPRS receives from the Max-Planck Society.
Work by Anton Vasyunin is supported by the Russian Ministry of Science and Higher Education via the State Assignment Contract FEUZ-2020-0038. Anton Vasyunin is a head of the Max Planck Partner Group at the Ural Federal University.
\end{acknowledgements}

\bibliographystyle{aa} 
\bibliography{Biblio} 

\begin{appendix}
\section{Layered ice:  Comparison of measurements and computational addition}
\label{compar_l_meas_comp}
\begin{figure*}[hbt!]
    \centering
    \vspace{0.5cm}
    \subfloat{\includegraphics[scale=.5]{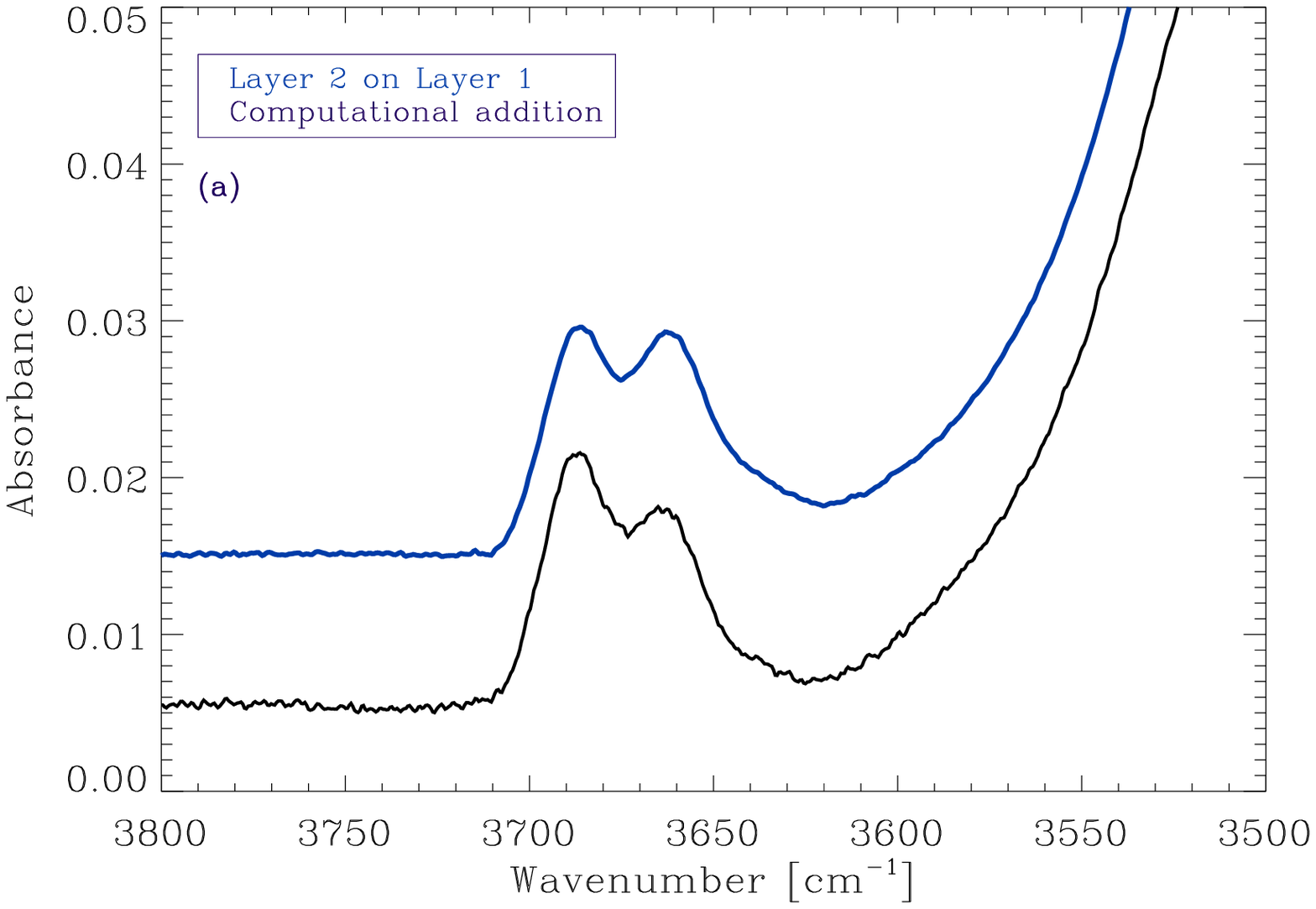}}
    \subfloat{\includegraphics[scale=.5]{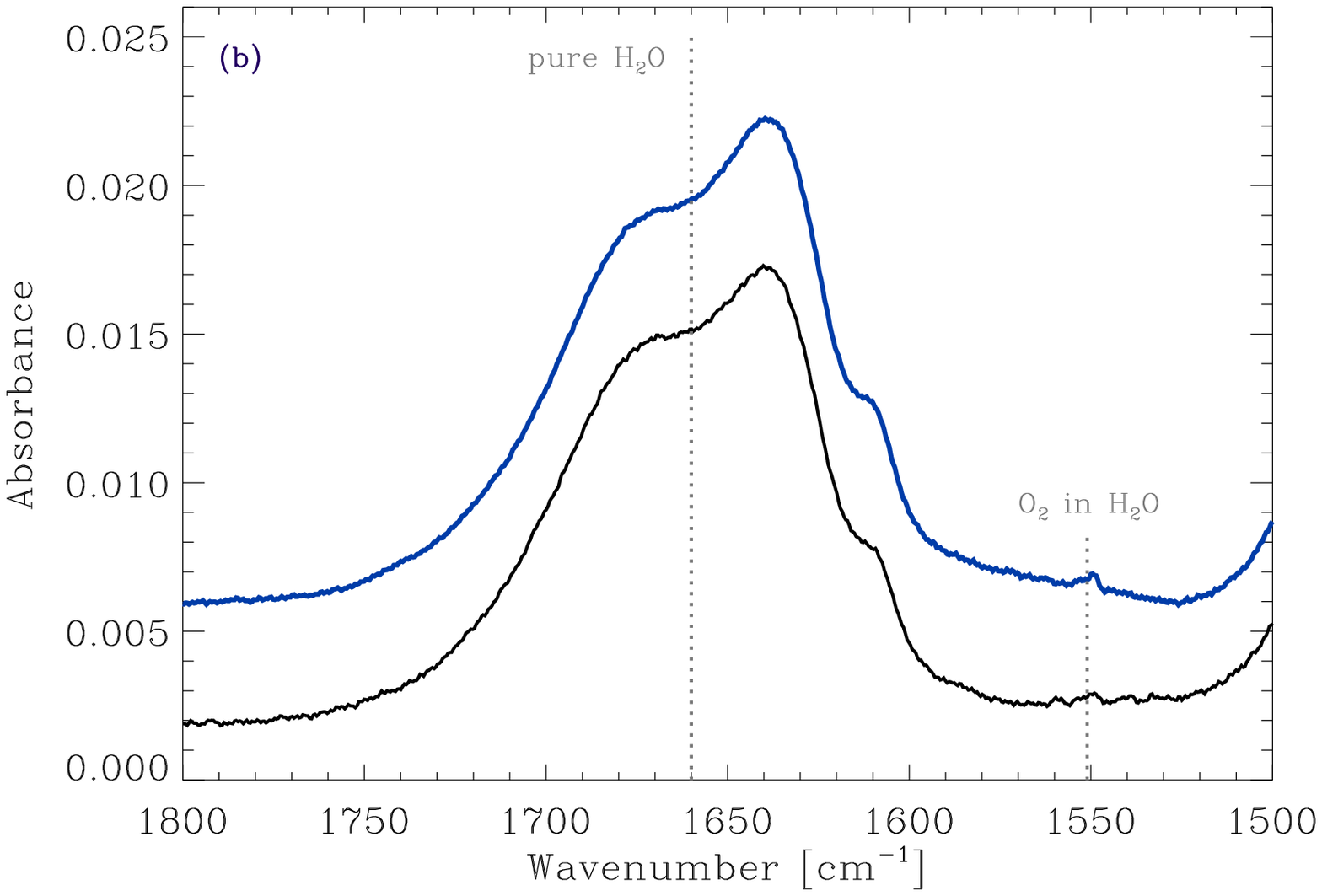}}
    
    \subfloat{\includegraphics[scale=.5]{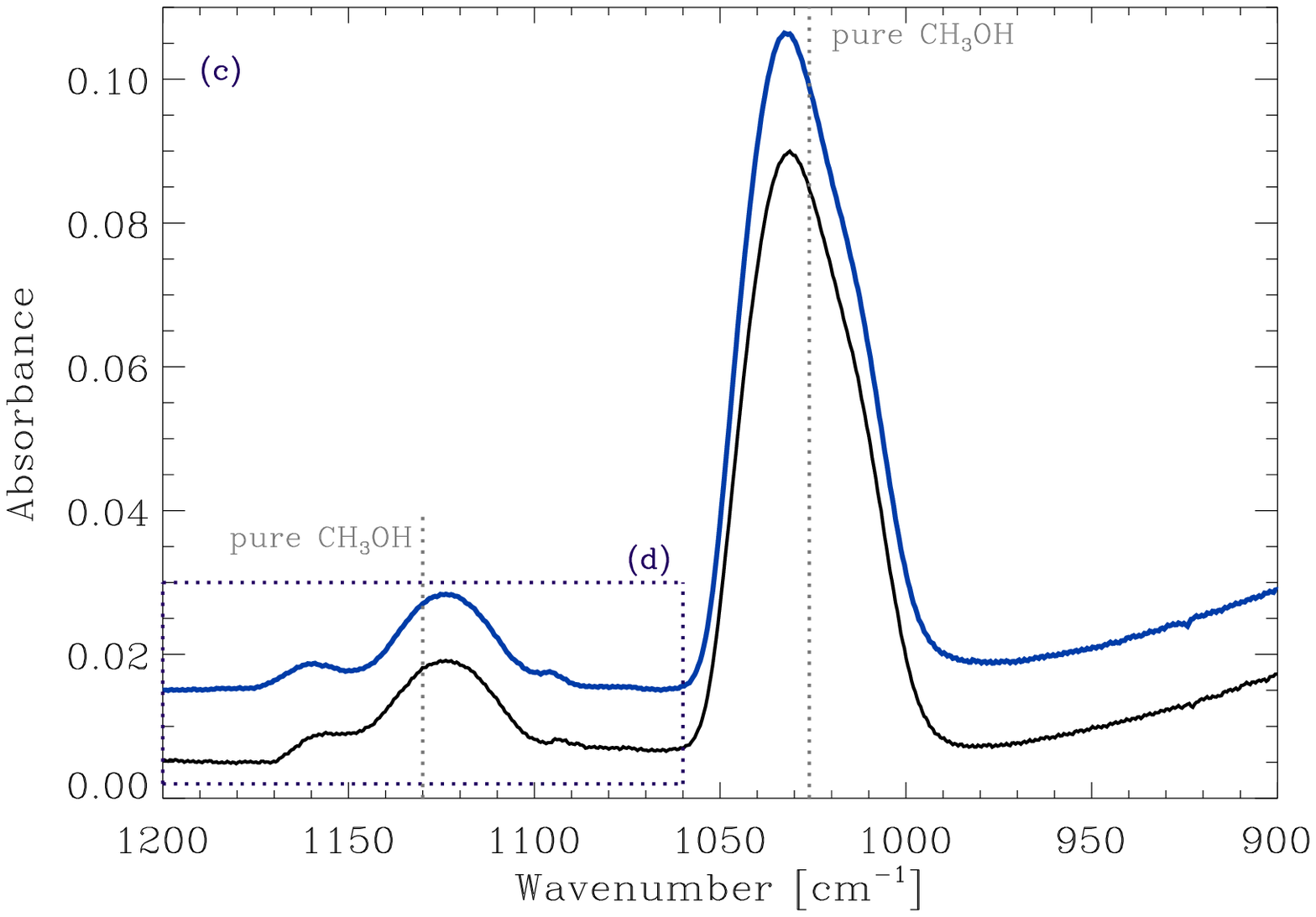}}
    \subfloat{\includegraphics[scale=.5]{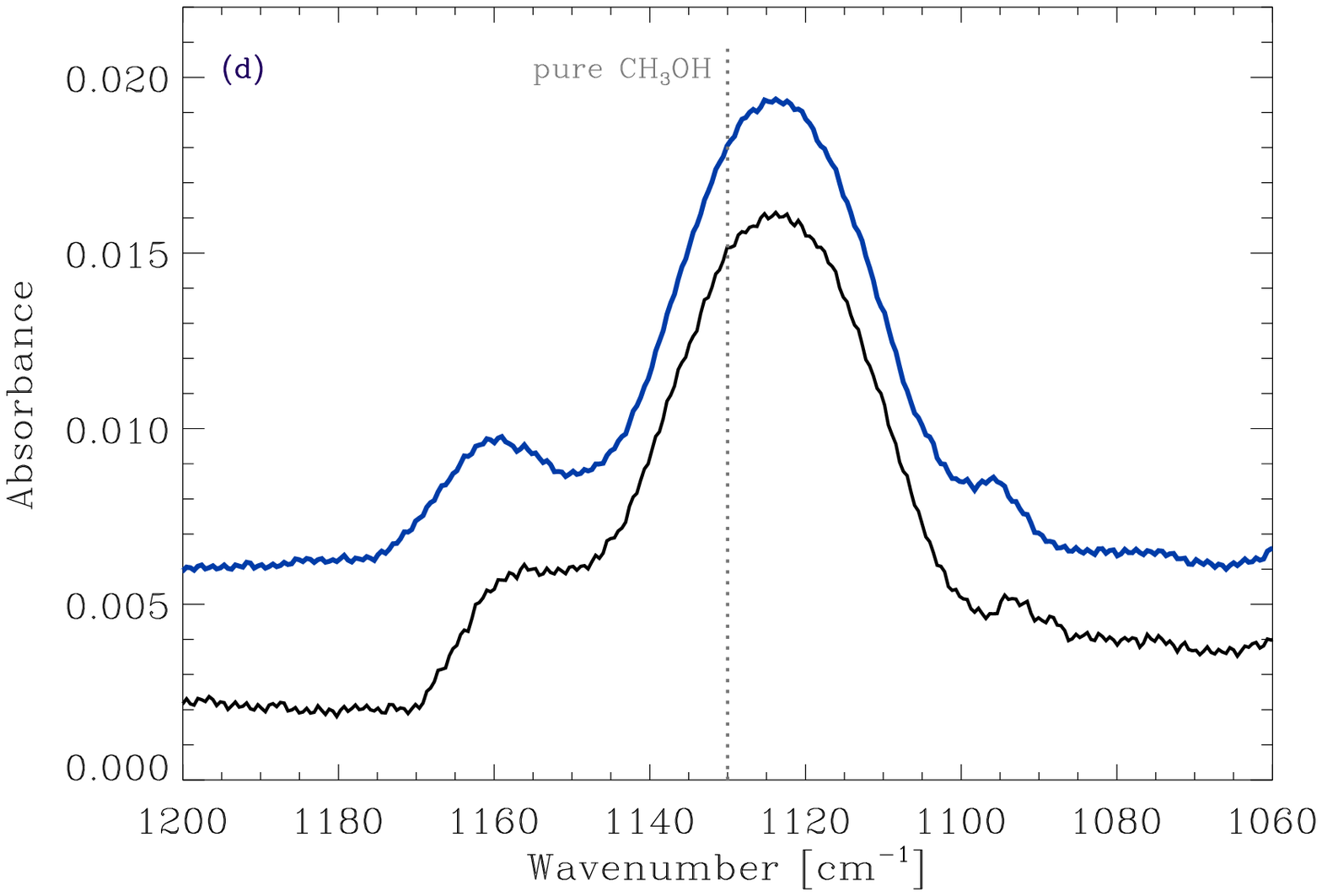}}
    \caption{Comparison of experimentally added (dark blue) and computationally added (black) layered ices, zoomed-in around the \ce{H2O} dangling bond (a), \ce{H2O} OH-bend (b), CO-stretch (c) and \ce{CH3}-rock (d) bands. Corresponding to Table \ref{mod_and_rel_water}, the spectra are shown for relative abundances of \ce{H2O}:\ce{CH3OH}:\ce{N2} = 100\%:67.5\%:27.5\% (layer 1) and \ce{H2O}:\ce{N2}:CO:\ce{O2} = 100\%:46\%:146\%:38\% (layer 2).}
    \label{layer_meas_comp}
\end{figure*}

\indent\indent During the analysis of the ice mixtures used for the layered experiments, we found some discrepancies in the position and shape of some absorption bands when we compared the spectra of layer 1 and layer 2 with the spectral features of the layered ice where layer 2 was deposited on top of layer 1.
While most of the absorption bands show no differences in the layered ice and the computational addition of the vibrational modes observed in layer 1 and 2, this is not the case for the \ce{H2O} dangling bonds, the \ce{O2} band, and the 1130~cm$^{-1}$ \ce{CH3} rocking band. The differences are shown in Fig. \ref{layer_meas_comp}.

When we compare the computational addition of layer 1 and 2 with the experimentally layered ice, we see for the dangling bonds that the intensity of the right local maximum at 3662~cm$^{-1}$ is remarkably weaker than the left local maximum. As we already observed a strong dependence of the 3662~cm$^{-1}$ maximum on the presence of CO and \ce{O2} the high abundance of these two molecules can explain the difference in the relative intensity of the band maxima via the interaction of the freely vibrating OH bonds with CO and \ce{O2} at the interface where layer 2 was deposited on top of layer 1.

This effect can also be seen in the increased band strength of \ce{O2} measured for the layered structure in comparison to the weaker band of the computationally added spectrum. The presence of the two polar species of layer 1, water and methanol, at the interface seems to amplify the intensity of the \ce{O2} band.

While it is true that increasing amounts of CO relative to \ce{CH3OH} cause a drop in the intensity of the 1130~cm$^{-1}$ shoulder features, as seen in \citet{Mueller2021}, spectroscopic results from the same work also show that the presence of small amounts of CO enhance these features when \ce{CH3OH} is layered on top of \ce{H2O}. This effect can be seen when experimentally and computationally added layered spectra are compared.
Moreover we see that the shoulders are shifted towards lower frequencies in the experimental data, whereas the maximum band position does not experience any effect due to the molecular interaction at the interface.
\end{appendix}

\end{document}